\documentclass[useAMS,usenatbib,usegraphicx]{mn2e}
\usepackage{fancyheadings}
\usepackage[dvips]{graphics}
\usepackage{colordvi}
\usepackage{rotating}
\usepackage{ifthen}
\usepackage{amssymb}
\usepackage{epsfig}
\usepackage{wrapfig}

\title[Spectro-astrometry  of Herbig Ae/Be stars]{On the binarity of Herbig Ae/Be stars}

\author[Deborah Baines, Ren\'e  Oudmaijer, John  Porter, Monica Pozzo] 
{Deborah Baines$^{1,2}$, Ren\'e D. Oudmaijer$^{2}$, John M. Porter$^{3, \dagger}$\thanks{It is with great sadness that we  have to report that,  when this paper was in its final stages,  John Porter suddenly died.}
, Monica Pozzo$^{4}$
 \\ 
$^{1}$ U.K. Gemini Support Group, Dept. of Astrophysics, Oxford
University, Keble Road, Oxford OX1 3RH, UK\\
$^{2}$ School of Physics and Astronomy, EC Stoner Building, University
of Leeds, Leeds LS2 9JT, UK\\
$^{3}$ Astrophysics Research Institute, Liverpool John Moores 
University, Twelve Quays House, Egerton Wharf, Birkenhead CH41 1LD,
UK \\
$^{4}$ Astrophysics Group, Imperial College London, Blackett
Laboratory, Prince Consort Road, London, SW7 2AZ, UK
}

\begin{document}

\date{Accepted --. Received --; in original form --}

\pagerange{\pageref{firstpage}--\pageref{lastpage}} \pubyear{2004}

\maketitle

\label{firstpage}

\begin{abstract} We present high resolution spectro-astrometry of a sample of 
28 Herbig Ae/Be and 3 F-type pre-main sequence stars. The
spectro-astrometry, which is essentially the study of unresolved
features in long-slit spectra, is shown from both empirical and
simulated data to be capable of detecting binary companions that are
fainter by up to 6 magnitudes at separations larger than $\sim$ 0.1
arcsec. The nine targets that were previously known to be a binary are
all detected.  In addition, we report the discovery of 6 new binaries
and present 5 further possible binaries. The resulting binary fraction
is 68$\pm$11\%. This overall binary fraction is the largest reported
for any observed sample of Herbig Ae/Be stars, presumably because of
the exquisite sensitivity of spectro-astrometry for detecting binary
systems. The data hint that the binary frequency of the Herbig Be
stars is larger than for the Herbig Ae stars. The appendix presents
model simulations to assess the capabilities of spectro-astrometry and
reinforces the empirical findings.  Most spectro-astrometric
signatures in this sample of Herbig Ae/Be stars can be explained by
the presence of a binary system. Two objects, HD~87643 and Z CMa,
display evidence for asymmetric outflows. Finally, the position angles
of the binary systems have been compared with available orientations
of the circumprimary disc and these appear to be co-planar. The
alignment between the circumprimary discs and the binary systems
strongly suggests that the formation of binaries with intermediate
mass primaries is due to fragmentation as the alternative, stellar
capture, does not naturally predict aligned discs.  The aligment
extends to the most massive B-type stars in our sample. This leads us
to conclude that formation mechanisms that do result in massive stars,
but predict random angles beween the binaries and the circumprimary
disks, such as stellar collisions, are also ruled out for the same
reason.
\end{abstract}


\section{Introduction}
In contrast to the low mass stars, which are commonly accepted to form
due to the gravitational collapse of a dusty cloud and the subsequent
magnetically controlled accretion via a disc (e.g. Bertout 1989), the
formation of the most massive stars is still very much uncertain. Part
of this is due to the extreme rarity of these objects and the fact
that they are very embedded in their molecular clouds, rendering them
obscured. To make headway, it is important to study the optically
visible Herbig Ae/Be stars as these have masses intermediate between
the two above cases.  The Herbig Ae/Be stars are now also found to
mark the transition between the different formation mechanisms of low
and high mass stars. Recent studies indicate that the Herbig Ae
stars present similar phenomena as T Tauri stars (Vink et al. 2003,
2005, see also Hubrig, Sch\"oller \& Yudin 2004), whereas the Herbig
Be stars are similar to the more embedded Massive Young Stellar
Objects (MYSOs, e.g. Drew et al. 1997).

To learn more about their formation, it is necessary to probe the
(circum-)stellar environments of Herbig Ae/Be stars at the small
scales where many important features for the further evolution such as
discs, outflows, and binaries are found. Millimetre observations have
detected discs in the size range of $\sim$ 200 to 3000 AU around
Herbig Ae/Be stars (Mannings \& Sargent, 1997 \& 2000; Pi\'{e}tu,
Dutrey \& Kahane 2003), while near-IR coronagraphic observations of
the Herbig Ae star HD 150193 (Fukagawa et al., 2003) and the Herbig Be
star HD 100546 (Grady et al., 2001) reveal discs extending to $\sim$
190 and 515 AU respectively. Using spectropolarimetry, Oudmaijer \&
Drew (1999) and Vink et al. (2002) found evidence for even smaller
scale discs, with sizes of the order of several stellar radii, around
Herbig Ae/Be stars.

Pre-main sequence binary fractions have been determined from IR
imaging (Bouvier \& Corporon, 2001; Leinert, Richichi \& Haas 1997;
Pirzkal, Spillar \& Dyck 1997; and Li et al., 1994) and optical
spectroscopy (Corporon \& Lagrange, 1999).  These studies are based on
limited samples of Herbig Ae/Be stars, but from extrapolations of the
detections, the inferred binary frequency is almost double that of
solar type Main Sequence stars for the period range probed (Leinert et
al. 1997), and  of the same order as the T Tauri stars
(K\"{o}hler \& Leinert, 1998).

In this paper we apply the technique of spectro-astrometry to study
Herbig Ae/Be stars. This method is particularly suited to detect
binary companions of emission line stars, and as we will show, can
detect secondaries as much as 6 magnitudes fainter than the
primary. As such it is among the most powerful methods to investigate binary
frequencies, while it has the potential to probe even closer to the
star and detect small scale structures in the circumstellar material.

The basic principle of spectro-astrometry is to measure the relative
spatial position of spectral features from a longslit spectrum
(Bailey, 1998). This is achieved by obtaining high spectral resolution
long-slit spectra of an object and determining the centre of the point
spread function (PSF) at each wavelength. In his pioneering paper,
Bailey (1998) already illustrated that spectro-astrometry is effective
at finding binaries and outflow structures around Herbig Ae/Be stars,
focusing on the strong H$\alpha$ emission of these objects. Later
studies have found outflows and binary signatures from a sample of T
Tauri stars (e.g. Takami, Bailey \& Chrysostomou 2003 and references
therein) and an outflow from a  Brown Dwarf (Whelan et al.  2005).

Returning to the detection of binary stars, a system where only one of
the components emits H$\alpha$ displays a simple and easily detectable
signature in the photo-centre of the spectrum. As the spatial profile
is the sum of both stars, the peak is not located at the position of
either star, but somewhere between the components, depending on the
intensity ratio - analogous to the centre of motion of binaries.
Because the H$\alpha$ line intensity ratio changes across the line,
the peak position shifts towards the H$\alpha$ dominant star and the
binary is revealed by an excursion in the positional spectrum.  We
also expect to see a decrease in the full-width-at-half-maximum (FWHM)
of the spatial profile across H$\alpha$ as it is dominated by only one
star. This paper is the first to present and use this additional
information that can be straightforwardly retrieved from the data.  We
will demonstrate later that the FWHM spectra have a strong diagnostic
potential in assessing the properties of the objects under
investigation.

In principle the method can be applied across any spectral line in
which the intensity ratios of the two stars change from the continuum
ratio; in the case of an absorption line the peak position will move
away from the star displaying the absorption.  Previously we reported on
data of the pre-main sequence binary HK Ori (Baines et al. 2004), the
aim of the present work is to further develop the technique and apply
it to a large sample of Herbig Ae/Be stars.  We observed 28 young
Herbig Ae/Be stars and 3 F-type pre-main sequence stars
spectro-astrometrically.  The observations are centred on H$\alpha$
because the H$\alpha$ emission line is usually by far the strongest
feature in the spectrum of these objects. It is therefore particularly
suitable to study the geometry of the circumstellar material, or as in
this paper, binarity.

In Section \ref{Observations} we describe the observations and the
targets. The method used to extract information from such data is
outlined here. Section \ref{Results4} contains our results, which is
followed by a discussion on our findings in Section 4. Section 5
summarizes the main conclusions and in the Appendix we discuss the
strengths and limits of spectro-astrometry by simulating data.

\begin{sidewaystable*}
 \centering

 \begin{minipage}{240mm}
{{\bf Table 1.} Log of the observations. Columns 2 to 4 list the RA,
Dec and spectral type of the objects respectively. The $V$ magnitudes
in column 5 are taken from SIMBAD. Column 6 lists the dates of the
observations and column 7 the exposure times. The average seeing for
each object is given in column 8 and the number of photons per pixel
in the stellar continuum in column 9. The 1$\sigma$ detection limit of
both the position spectra and FWHM spectra are given in columns 10 and
11 respectively. These are the root-mean-square variations measured in
line-free regions in the stellar continuum. Properties of the lines
are listed in the final columns. The emission line profiles are classified
as doublepeaked (D), singly peaked (S) or as showing (inverse) P Cygni (P)
profiles. The equivalent width, with estimated errors of 10\%, is also
listed.  References: for the spectral types: 1. Mora et. al. (2001),
2. SIMBAD, 3. Th\'e et. al. (1994).
\label{log} }

  \begin{tabular}{lcclrccrrrrrrrcc}
  \hline
Object    &     RA     &     DEC     &     Spec.                    &{\it V}&  Date    & t$_{exp}$& Seeing & Photons &    rms$_{pos}$ &     rms$_{FWHM}$ &     H$\alpha$ W$_\lambda$ &     H$\alpha$ \\
          &     (2000) &     (2000)  &     type                      &      &  &    (s) &     (arcsec) &     (x10$^4$) &     (mas) &     (mas) &     ($\rm \AA$) &     Profile \\

\hline
XY Per    & 03 49 36.3 & +38 58 55.5 & A2{\sc i}{\sc i}$^2$      & 9.4  & 21-09-02 & 6 x 3000 & 2.2 &  6.7 & 2.0 &  4.3 & -6.6 & D   \\
 ''       & ''         & ''          & ''                        & ''   & 23-09-02 & 8 x 900  & 2.1 &  6.1 & 4.6 & 21.0 & -4.5 & D   \\ 
AB Aur    & 04 55 45.8 & +30 33 04.3 & A0$^2$                    & 7.1  & 29-01-02 & 4 x 180  & 2.4 & 16.0 & 2.5 &  4.5 & -45  & P   \\
 ''       &  ''        & ''          & ''                        & ''   & 19-09-02 & 8 x 510  & 1.8 & 32.0 & 0.9 &  1.7 & -21  & P   \\
HD 31648  & 04 58 46.3 & +29 50 37.0 & A5{\sc v}$^1$             & 7.7  & 29-01-02 & 4 x 300  & 2.6 & 16.5 & 2.3 &  5.0 & -17  & D   \\  
 ''       & ''         & ''          &  ''                       & ''   & 19-09-02 & 4 x 1440 & 1.6 & 27.0 & 1.1 &  2.7 & -14  & D/P \\
HK Ori    & 05 31 28.0 & +12 09 10.3 & A4/G1{\sc v}$^1$          & 11.9 & 29-01-02 & 8 x 300  & 2.4 &  1.6 & 7.7 & 19.2 & -47  & D   \\
HD 244604 & 05 31 57.3 & +11 17 41   & A3$^2$                    & 9.4  & 28-01-02 & 4 x 600  & 1.9 & 10.5 & 2.1 &  4.5 & -1.7 & D   \\
T Ori     & 05 35 50.4 & -05 28 35   & A3{\sc i}{\sc v}$^1$      & 9.5  & 28-01-02 & 4 x 600  & 1.6 &  6.7 & 2.1 &  5.3 & -8   & D   \\
 ''       & ''         & ''          & ''                        & ''   & 29-01-02 & 4 x 600  & 2.0 &  4.9 & 2.6 &  6.5 & -8   & D   \\
CQ Tau    & 05 35 58.5 & +24 44 54.1 & F5{\sc i}{\sc v}$^1$      & 10.7 & 23-09-02 & 4 x 2400 & 1.8 &  4.7 & 3.4 &  8.2 & -1.6 & invP\\
V380 Ori  & 05 36 25.4 & -06 42 57.7 & A0$^2$                    & 10.0 & 29-01-02 & 4 x 300  & 2.5 &  1.4 & 8.3 & 19.1 & -80  & S/D \\
MWC 137   & 06 18 45.5 & +15 16 52.4 & B0$^3$                    & 11.2 & 22-09-02 & 8 x 600  & 1.2 &  1.7 & 2.3 & 11.2 & -370 & S   \\
HD 45677  & 06 28 17.4 & -13 03 11.1 & B3$^3$                    & 8.0  & 29-01-02 & 8 x 300  & 2.8 & 13.8 & 2.8 &  6.8 & -57  & D   \\
MWC 147   & 06 33 05.2 & +10 19 20.0 & B6$^2$                    & 8.8  & 28-01-02 & 4 x 600  & 2.1 & 20.5 & 1.7 &  4.2 & -56  & D   \\
MWC 158   & 06 51 33.4 & -06 57 59.4 & B9$^2$                    & 6.6  & 28-01-02 & 8 x 45   & 1.6 & 22.0 & 1.3 &  3.2 &  -71 & D   \\
 ''       & ''         & ''          & ''                        & ''   & 20-09-02 & 24 x 120 & 1.4 & 36.0 & 0.9 &  1.7 &  -77 & D   \\
GU CMa    & 07 01 49.5 & -11 18 03.3 & B2{\sc v}$^2$             & 6.6  & 28-01-02 & 4 x 300  & 1.6 & 85.0 & 1.4 &  1.7 &  -7  & D   \\
Z CMa     & 07 03 43.2 & -11 33 06.2 & B5$^3$                    & 9.9  & 28-01-02 & 4 x 600  & 1.8 & 11.0 & 1.5 &  4.7 & -68  & D   \\
MWC 166   & 07 04 25.5 & -10 27 15.7 & B0{\sc i}{\sc v}$^2$      & 7.0  & 28-01-02 & 4 x 240  & 1.8 & 33.0 & 1.1 &  2.4 & -1   & D   \\
HD 58647  & 07 25 56.1 & -14 10 43.6 & B9{\sc i}{\sc v}$^2$      & 6.8  & 29-01-02 & 4 x 180  & 2.6 & 23.5 & 2.0 &  5.3 & -6.0 & D   \\
HD 85567  & 09 50 28.5 & -60 58 03.0 & B5{\sc v}$^3$             & 8.6  & 29-01-02 & 4 x 450  & 2.1 & 16.5 & 2.0 &  4.8 & -40  & S   \\
HD 87643  & 10 04 30.3 & -58 39 52.1 & B2$^2$                    & 8.9  & 28-01-02 & 8 x 120  & 1.2 & 10.0 & 1.2 &  3.0 & -145 & P,D \\
HD 95881  & 11 01 57.6 & -71 30 48.4 & A1/2{\sc iii}/{\sc iv}$^2$&8.3   & 29-01-02 & 4 x 300  & 1.9 & 15.0 & 2.0 &  3.9 & -11.5& D   \\  
HD 97048  & 11 08 03.3 & -77 39 17.5 & A0$^2$                    & 8.5  & 28-01-02 & 4 x 450  & 1.7 & 22.7 & 1.4 &  3.0 & -25  & S   \\
HD 98922  & 11 22 31.7 & -53 22 11.5 & B9{\sc v}$^2$             & 6.8  & 29-01-02 & 4 x 120  & 1.8 & 24.7 & 1.2 &  3.2 & -17.5& P   \\
HD 100546 & 11 33 25.4 & -70 11 41.2 & B9{\sc v}$^2$             & 6.7  & 28-01-02 & 4 x 60   & 1.4 & 15.3 & 1.4 &  3.0 & -35  & S   \\
HD 104237 & 12 00 05.1 & -78 11 34.6 & A4$^3$                    & 6.6  & 29-01-02 & 4 x 120  & 2.0 & 25.4 & 1.6 &  3.6 & -26  & D/P \\
HD 135344 & 15 15 48.4 & -37 09 16.0 & A0/F4{\sc v}$^3$          & 8.6  & 29-01-02 & 4 x 600  & 1.7 & 25.4 & 1.3 &  3.0 & -6.5 & S/D \\
HD 142527 & 15 56 41.9 & -42 19 23.3 & F6{\sc i}{\sc ii}$^2$     & 8.3  & 29-01-02 & 4 x 450  & 1.7 & 30.0 & 1.2 &  3.0 & -5.8 & P   \\ 
51 Oph    & 17 31 25.0 & -23 57 45.5 & B9.5{\sc iii}$^1$         & 4.8  & 29-01-02 & 4 x 30   & 2.0 & 28.0 & 1.4 &  3.5 & +4   & D   \\
HD 190073 & 20 03 02.5 & +05 44 16.7 & A2{\sc i}{\sc v}$^1$      & 7.8  & 21-09-02 & 8 x 1250 & 2.0 & 30.0 & 1.1 &  2.7 & -24  & P   \\
MWC 361   & 21 01 36.9 & +68 09 47.8 & B2{\sc v}$^2$             & 7.4  & 19-09-02 & 16 x 500 & 1.3 & 49.0 & 0.9 &  1.9 & -59  & D   \\
BH Cep    & 22 01 42.9 & +69 44 36.5 & F5{\sc i}{\sc v}$^2$      & 10.8 & 22-09-02 & 4 x 2700 & 1.5 &  3.6 & 2.7 &  7.0 & 0.0  & D/T \\
SV Cep    & 22 21 33.3 & +73 40 18   & A0$^2$                    & 10.1 & 22-09-02 & 4 x 1800 & 1.7 &  3.3 & 3.8 &  7.2 & +1.2 & D   \\
MWC 1080  & 23 17 26.1 & +60 50 43   & B0$^2$                    & 11.6 & 23-09-02 & 4 x 2400 & 1.8 &  2.1 & 4.5 &  8.7 & -140 & P/D \\
\hline
\end{tabular}
\hspace*{\fill}
\end{minipage}
\end{sidewaystable*}

\section{Observations and data reduction}\label{Observations}

The data presented here are due to two observing runs, one in the
southern and one in the northern hemisphere.  The Herbig Ae/Be stars
were selected from ``Table 1'' in the catalogue of Th\'{e}, de Winter
\& Perez (1994). All targets were chosen to be bright enough to obtain
high S/N ratios in reasonable exposure times, and no previous
knowledge of known binarity or otherwise was taken into account for
the selection of the stars. Three objects, V380 Ori, Z CMa and HD
97048, that were observed by Bailey (1998) using spectro-astrometry
were included to allow a consistency check.  As a significant
subset from Th\'e et al.'s catalogue was observed, the final sample is
a representative selection of the intermediate mass pre-main sequence
Herbig Ae/Be stars.  Thirty-one objects were observed, 15 have B
spectral type, 13 have spectral type A, while the remaining 3 objects
are of F type. The log of the observations is shown in Table
\ref{log}.

Observations were carried out on the 28th and 29th of January 2002 at
the 3.9 m Anglo-Australian Telescope (AAT) using the RGO spectrograph
with its 82 cm camera and a MITLL 2048 x 4096 CCD. The set-up gave a
spatial pixel size of 0.15 arcsec, providing very good sampling of the
seeing, which ranged from $\sim$ 1.2 to 2.6 arcsec. A 1200 line
mm$^{-1}$ grating was used to give a wavelength coverage of  $\sim$6300
to 6800 $\rm \AA$. A 1 arcsec slit width resulted in a spectral
resolution of about 40 kms$^{-1}$ at H$\alpha$ sampled with $\sim$
0.15 {\rm \AA} pixels. The spectra were obtained at four slit position
angles (PAs, at 0$^{\circ}$, 90$^{\circ}$, 180$^{\circ}$, and
270$^{\circ}$) for each object. This is to ensure optimal correction
for any systematic effects such as curvature or optical distortion
introduced by the spectrograph or a misalignment of the spectrum with
the CCD columns.  Wavelength calibrations were made from observations
of a CuAr lamp.

Further observations were carried out from the 19th to the 23rd of
September 2002 at the 2.5 m Isaac Newton Telescope, located at the
Roque de Los Muchachos Observatory, La Palma, Spain. The IDS
spectrograph with its 50 cm camera and an EEV10 2048 x 4096 CCD was
used, giving a spatial pixel size of 0.19 arcsec. The seeing varied
between $\sim$ 1.3 to 2.5 arcsec. The atmospheric conditions were
worse during the early hours of 22nd September due to the presence of
cirrus. An 1800 line mm$^{-1}$ grating was used to give the same
wavelength coverage as the AAT set-up, above. The 1 arcsec slit width
resulted in a resolution of $\sim$ 20 kms$^{-1}$ at H$\alpha$ (sampled
with $\sim$0.13 {\rm \AA} \, pixel$^{-1}$). The spectra were obtained, as
at the AAT, at four slit PAs (0$^{\circ}$, 90$^{\circ}$,
180$^{\circ}$, and 270$^{\circ}$) for each object, followed with an
arc spectrum of a CuNe lamp.

\setcounter{table}{1}
\begin{table*}
\begin{center}
\caption{The previously known binaries, the newly discovered binaries
and possible binaries. Columns 2 and 3 list the binary separations and
position angles from the literature, where available. Columns 4, 5 and
6 denote the position offset and the position angle of the binary
system and FWHM change over the H$\alpha$ profile from our
spectro-astrometric data.  Note that the positional shifts across the
line will always be {\it smaller} than the true separation (see
text). The ``possible'' binaries were classified as such because they
only show binary signatures in the FWHM spectra.  Superscripts for the
literature values are the references: 1. Fu et al (1997);
2. Leinert et al (1997); 3. Corporon (1998); 4. Pirzkal et al (1997);
5. Ghez, Neugebauer \& Matthews (1993); 6. ESA (1997).
\label{binarytable}}
\begin{tabular}{lllllllll}
\hline
 NAME    &   Lit Sep         &  Lit PA              &  H$\alpha$ shift &  PA           &  $\Delta$FWHM \\
         &  (arcsec)         &  (deg)               &  (arcsec)        &  (deg)        & (arcsec)            \\
\hline
Known Binaries:\\
\hline
GU CMa   & 0.655$\pm$0.007$^1$   & 192.6$\pm$0.5    & 0.132$\pm$0.005  & 192.5$\pm$0.2 & 0.14$\pm$0.01  \\
HK Ori   & 0.34$\pm$0.02$^2$     & 41.7$\pm$0.5     & 0.11$\pm$0.01    & 40$\pm$1      & -   \\
MWC 147  & 3.10$\pm$0.05$^3$     & -                & 0.007$\pm$0.002  & 72$\pm$3      & 0.03$\pm$0.01  \\ 
MWC 166  & 0.648$\pm$0.007$^1$   & 298.2$\pm$0.5    & 0.050$\pm$0.002  & 287.0$\pm$0.4 & 0.05$\pm$0.01  \\
MWC 361  & 2.25$\pm$0.24$^4$     & 164$\pm$1        & 0.005$\pm$0.001  & 5$\pm$3       & 0.005$\pm$0.003  \\
MWC 1080 & 0.76$\pm$0.02$^2$     &  267$\pm$1       & 0.272$\pm$0.004  & 267.3$\pm$0.3 & 0.27 $\pm$0.01   \\
XY Per   &  1.331$\pm$0.010$^6$  & 76.300$\pm$0.001 & 0.242$\pm$0.002  & 85.0$\pm$0.3  & 0.28$\pm$0.01  \\ 
"        &  "                    & "                & 0.335$\pm$0.006  & 83.8$\pm$0.3  & 0.25$\pm$0.03  \\   
V380 Ori & 0.154$\pm$0.002$^2$   & 204.2$\pm$0.9    & 0.04$\pm$0.01    & 204$\pm$3     & -   \\
Z CMa    & 0.100$\pm$0.007$^2$   & 122.5$\pm$0.2    & 0.041$\pm$0.002  & -             & 0.06$\pm$0.01  \\
\hline
New Binaries: \\
\hline
AB Aur   &                   &                      & 0.029$\pm$0.003  & 163$\pm$6     & 0.12$\pm$0.01   \\
"        &                   &                      & 0.023$\pm$0.001  & 129$\pm$2     & 0.07$\pm$0.01   \\
HD 45677 &                   &                      & 0.013$\pm$0.004  & 150$\pm$17    & 0.05$\pm$0.01   \\
HD 58647 &                   &                      & 0.016$\pm$0.003  & 115$\pm$10    & 0.03$\pm$0.01   \\
HD 85567 &                   &                      & 0.029$\pm$0.003  & 1$\pm$2       & 0.05$\pm$0.01   \\
HD 98922 &                   &                      & 0.039$\pm$0.002  & 0$\pm$1       & 0.010$\pm$0.005   \\
MWC 158  &                   &                      & 0.016$\pm$0.002  & 30$\pm$7      & 0.050$\pm$0.005   \\
\hline
Possible Binaries : \\
\hline
HD 95881  &                  &                      &                  &                   &  0.020$\pm$0.006  \\ 
HD 104237 &                  &                      &                  &                   &  0.04$\pm$0.01  \\ 
HD 142527 &                  &                      &                  &                   &  0.04$\pm$0.01  \\
HD 190073 &                  &                      &                  &                   &  0.020$\pm$0.007  \\ 
HD 244604 &                  &                      &                  &                   &  0.02$\pm$0.01  \\ 
\hline
\end{tabular}
\end{center}
\end{table*}

The data were reduced using the IRAF package (http://iraf.noao.edu).
High signal-to-noise flat fields were made by combining many exposures
with the spectrograph illuminated by a tungsten lamp. After
subtracting the bias level and dividing by a normalised flat field,
the 2-dimensional spectrum was fitted by Gaussian profiles in the
spatial direction at each wavelength step. Visual inspection of the
data confirmed that Gaussians are a proper representation of the real
profiles. This results in a so-called position spectrum and FWHM
spectrum, which gives the centre of the emission and FWHM of the
emission as a function of wavelength respectively. Four position and
four FWHM spectra were obtained at the different position angles. Any
instrumental effects are largely eliminated by averaging those with
opposite position angles (0$^{\circ}$-180$^{\circ}$, or
90$^{\circ}$-270$^{\circ}$). This then returns two position and two
FWHM spectra, one in the North-South (NS) direction and the other in
the East-West (EW) direction.

The root-mean-square variations which give a measure of the
uncertainty in the positioning of the centroid and FWHM were on average
2.2 mas and 5.5 mas respectively. These values were measured in a
seeing which varied from 1.2 to 2.8 arcsec, and signal-to-noise ratios
in the continuum ranging from 120 to 920 (see Table \ref{log}). The
theoretical precision of the centroid position is better for smaller
seeing and higher SNR (Bailey, 1998; Porter, Oudmaijer \& Baines
2004), and can easily reach sub-pixel values.

As Bailey (1998) noted, artifacts in the position spectrum can arise
when sharp unresolved lines are observed. He suggested that such
features are caused by telescope tracking errors or a consequence of
seeing. One of the ways to eliminate such instrumental artifacts is to
observe the objects at four different position angles, any real
features should appear as mirror images of each other when obtained
with slit Pa's 180$^{\circ}$ apart, while instrumental effects only
appear in one direction.

In reality, this only reduces the effects to a certain extent. This is
especially the case when steep changes in the flux spectrum occur, and
visual inspection of the data is necessary to assess the reality of
any resulting features.  This is an important issue because many of
the target stars have a strong double peaked H$\alpha$ profile with a
dramatic change in intensity from the double peaks to the central
absorption dip. Although the central absorption dips themselves are
resolved, they still produce the type of signature in the position
spectrum described by Bailey.  Similarly, any non-uniformity in pixel
to pixel response can produce features which are no true signatures in
the stellar spectra. Despite careful flatfielding, there is a number
of pixels where this effect has not been fully eliminated. Any spectra
falling across these coordinates show the same displacement.  All
systematic effects that have been identified are indicated by dotted
lines in the spectro-astrometric plots shown later.

\begin{figure*}
\centering
\epsfig{file=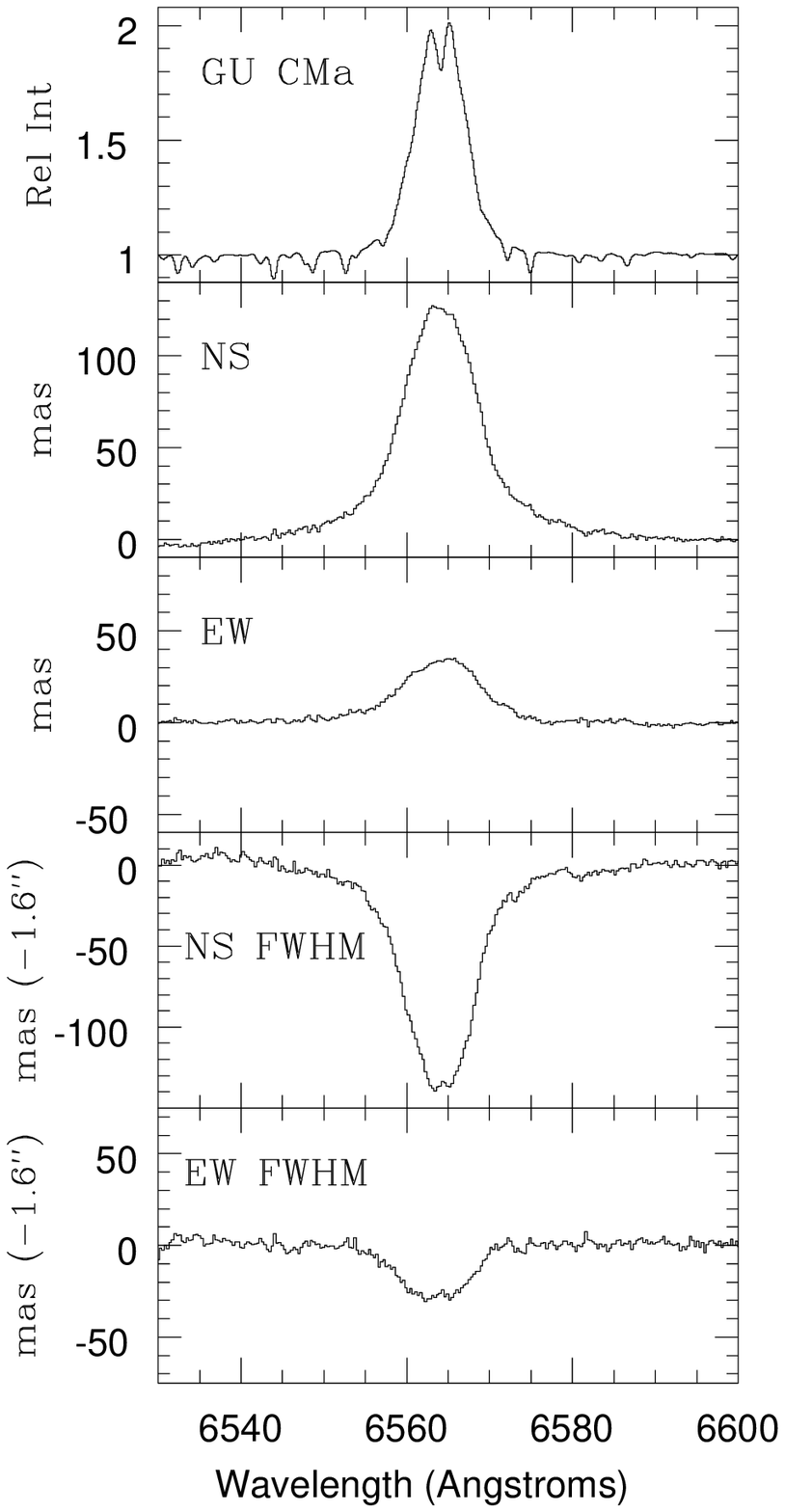,width=5.75cm}
\epsfig{file=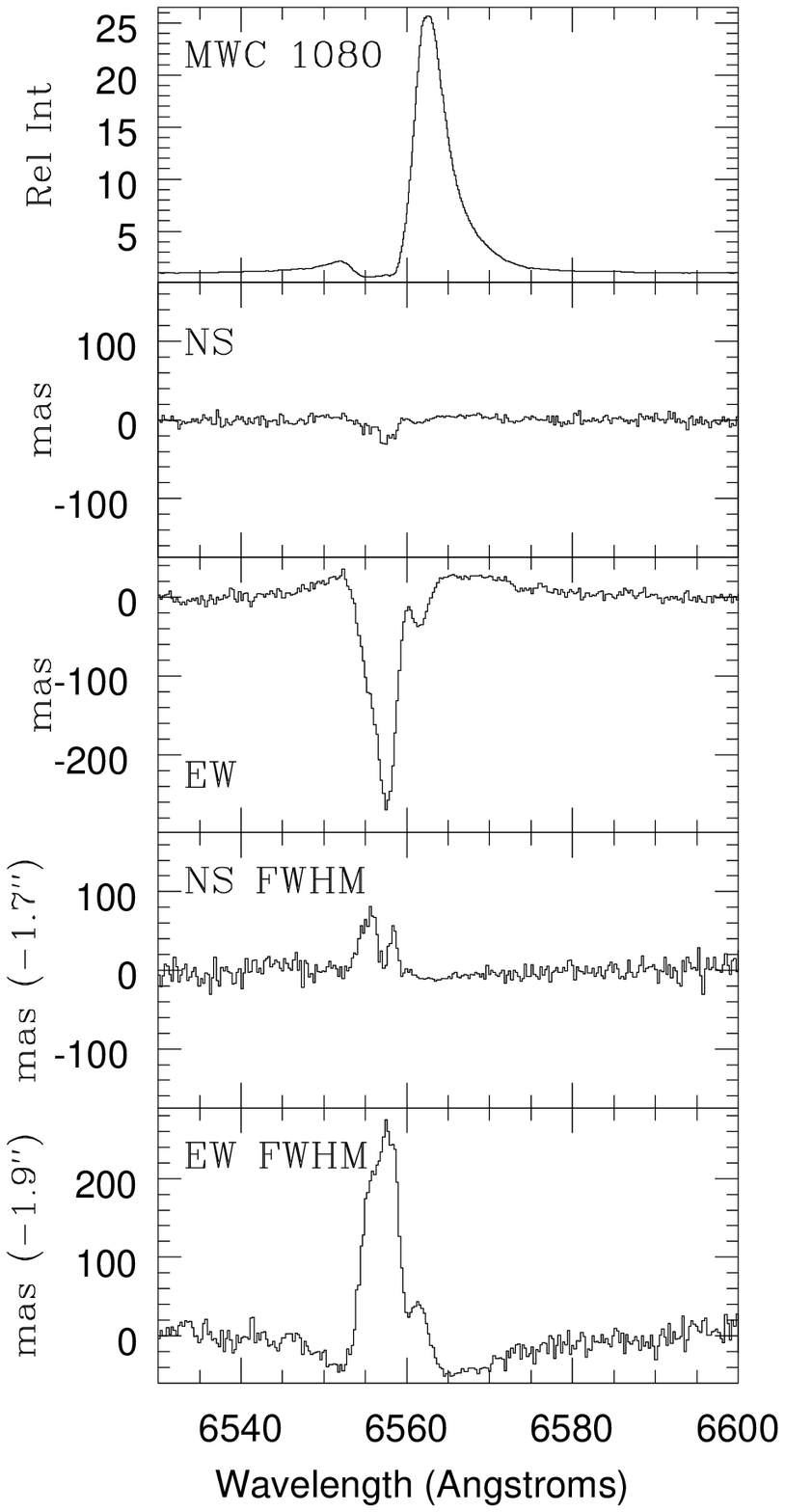,width=5.75cm}
\caption{The spectro-astrometric plots of GU CMa (left) and MWC 1080
(right) centred on H$\alpha$. The graph shows, from top to bottom, the
intensity spectrum, the centroid position of the spectrum in the
North-South and East-West direction (with North and East up) and the
FWHM of the spatial profiles of the NS and EW spectra respectively.
For the FWHM spectra the vertical axis label includes the average
seeing during the respective observations.  
Each position and FWHM spectrum has an arbitrary zero-point,
which is adjusted to correspond to the continuum position and
continuum FWHM position respectively.
 \label{GUCMa}}
\end{figure*}

\section{Results}\label{Results4}

Out of 31 targets, we detect spectro-astrometric excursions in more
than 20. The effects can often be simply explained with the presence
of a binary companion, the results are summarized in Table~2. In the
following we will start discussing the known binaries in the
sample. As we know their system parameters, they provide an important
check on the method. We then discuss the new and possible binaries,
while the 2 objects with evidence for asymmetric outflows are
discussed separately.

\subsection{Binaries}\label{binaries}

We begin our description by presenting GU CMa and MWC~1080. These two
objects illustrate different aspects in spectro-astrometric data of
binary systems.

\smallskip

{\it GU CMa :} GU CMa is a B2{\sc v} Herbig Ae/Be binary with a
separation of 0.655 $\pm$ 0.007 arcsec and a PA of 192.6 $\pm$
0.5$^{\circ}$ (Fu et al., 1997).  The North-Eastern star dominates the
optical light.  Figure \ref{GUCMa} displays the data of the
star. Large positional displacements across H$\alpha$ towards the
North and East are present, indicating that the optically dominant
object also dominates the H$\alpha$ emission. The excursions are a
lower limit to the true separation as in the continuum both stars
contribute to the spectrum.  The FWHM spectrum changes across the
H$\alpha$ line, as expected for a binary with one component dominating
the H$\alpha$ emission. The FWHM in the continuum measures the
unresolved binary, and decreases across H$\alpha$ when only the
H$\alpha$ emitting component is measured.  From the excursion of the
H$\alpha$ line we derive the position angle of the binary to be 192.5
$\pm$ 0.2$^{\circ}$. This is very close to the PA found by Fu et al
(1997).

\smallskip

{\it MWC 1080 : } A different effect is visible when we consider an
object which displays an H$\alpha$ P Cygni profile.
MWC 1080 is a B0 Herbig Ae/Be star and Leinert et al (1997) measured a
separation of 0.76 $\pm$ 0.02 arcsec and a PA = 267 $\pm$ 1$^{\circ}$.
Fig.~\ref{GUCMa} shows the spectro-astrometric data. H$\alpha$ is
extremely strong with a peak that is 25 times the continuum, and a
deep P Cygni absorption.  In contrast to GU CMa, the positional
spectrum exhibits more than one feature.  Across the H$\alpha$
emission there is a small displacement to the East, coinciding with a
decrease in the FWHM. As discussed above, such displacements are
expected from a binary system with H$\alpha$ emission from one of the
components.

The different aspect of MWC 1080 is the large westwards displacement
across the H$\alpha$ P Cygni {\it absorption}. This is accompanied by
a large increase in the FWHM and can be straightforwardly explained
by the binary nature of the system. The relative contribution
of the secondary increases as the light from the H$\alpha$ emitting
primary diminishes across the P Cygni absorption. The centroidal
position shifts to the secondary, while the FWHM now measures both
stars and  increases.  From the positional excursions, we
measure a PA of 267.3 $\pm$ 0.3$^{\circ}$, very close to that derived
from the imaging data reported by Leinert et al. (1997).

\smallskip

GU CMa and MWC 1080 are good examples of the technique of
spectro-astrometry. Both objects were observed in seeing between 1.5
and 2 arcsec, but despite the fact that their separations are much
smaller than the seeing, they were  clearly and unambiguously
revealed as binary objects. The introduction of the FWHM spectrum
proves to be particularly useful in assessing whether any positional
displacements are due to the presence of a binary.  The excursions
towards the H$\alpha$ emitting star allowed us to derive the position
angle of the binary system. The resulting values are extremely close
to those measured from imaging.

We continue our discussion with the full sample of the known binaries
(with existing data listed in Table~\ref{binarytable}), and then
further discuss the sample as a whole.

\subsubsection{Known binaries}\label{KnownBins}

\begin{figure*}
\centering
\epsfig{file=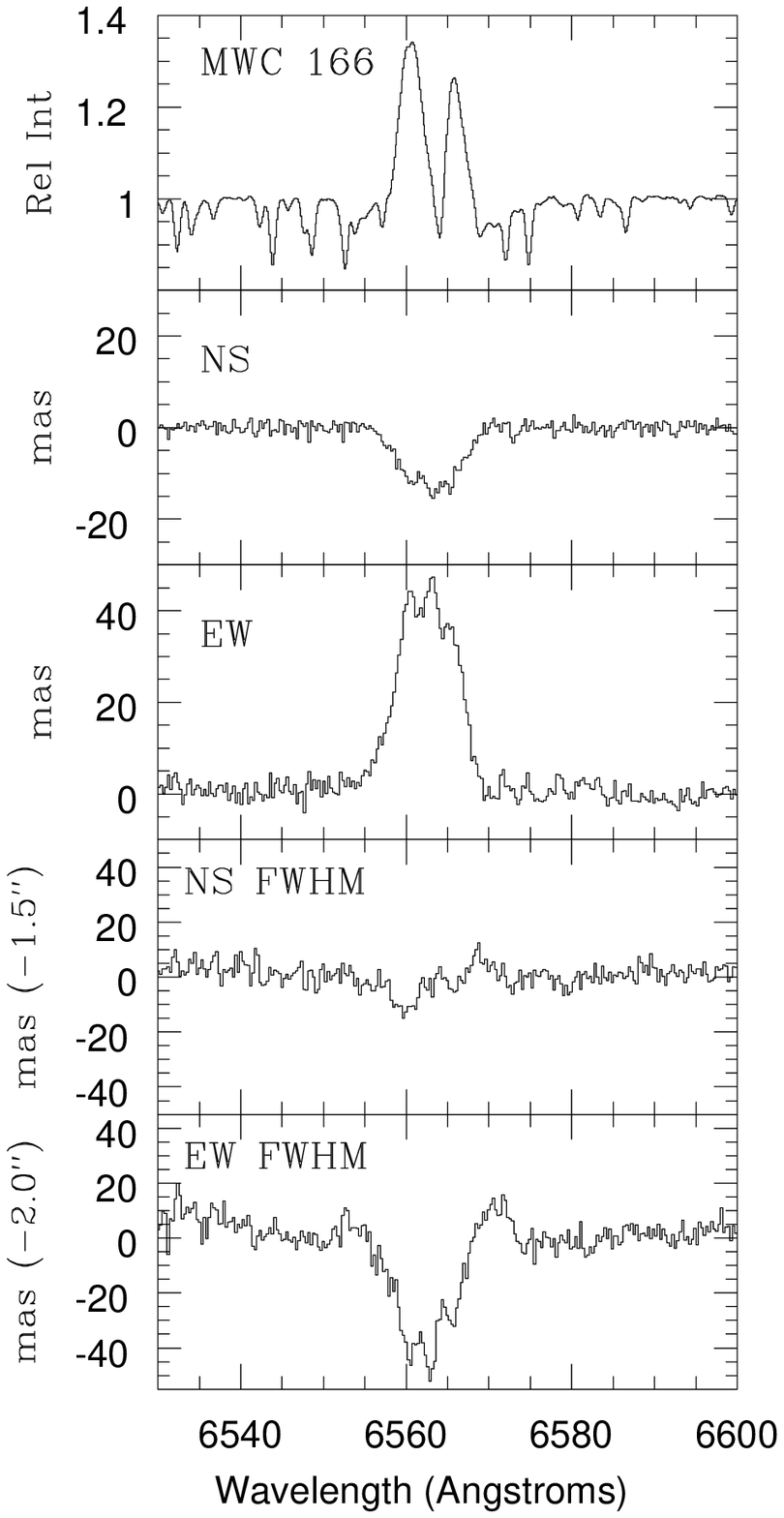,width=6.0cm}
\epsfig{file=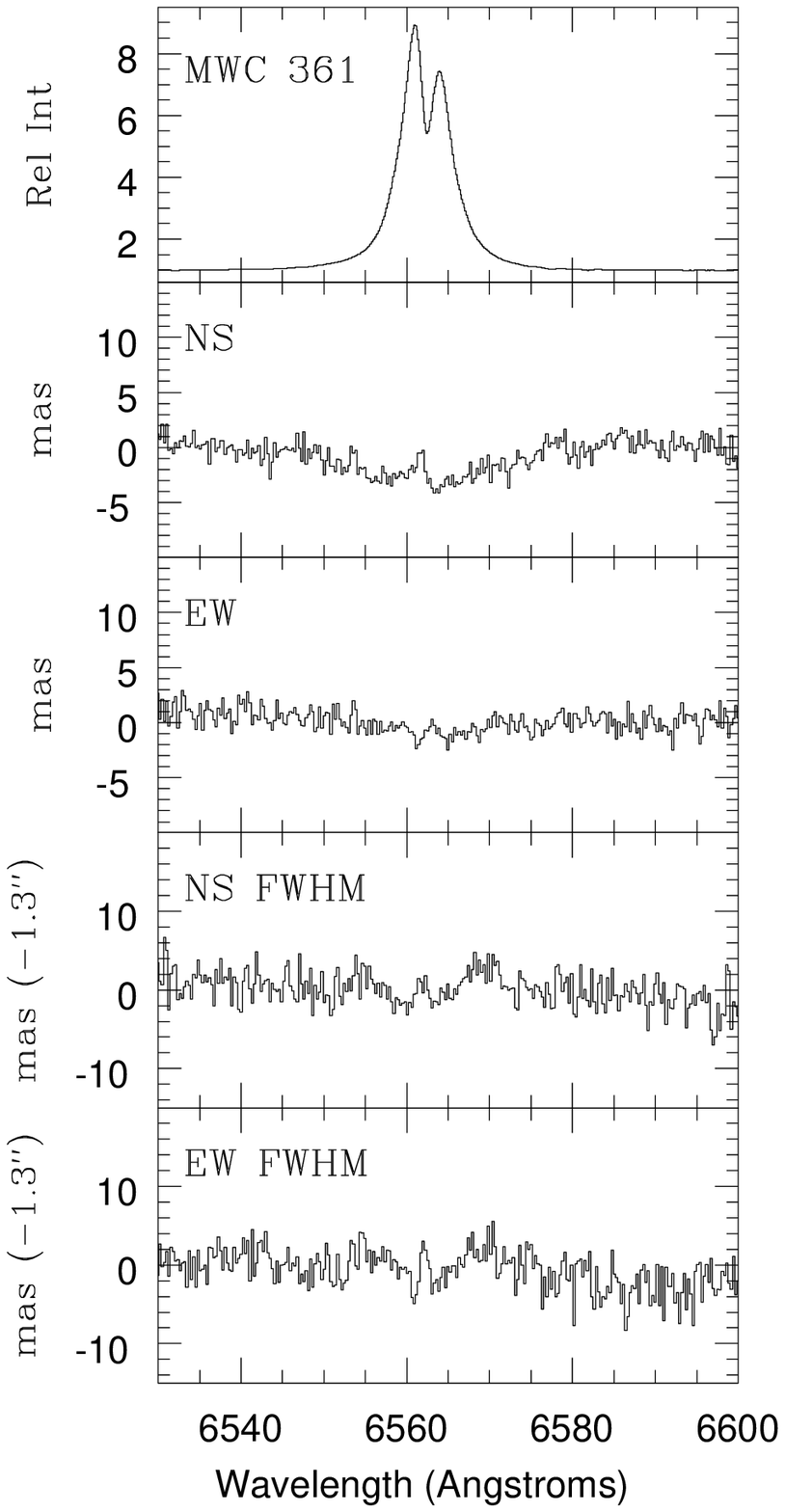,width=6.0cm}
\epsfig{file=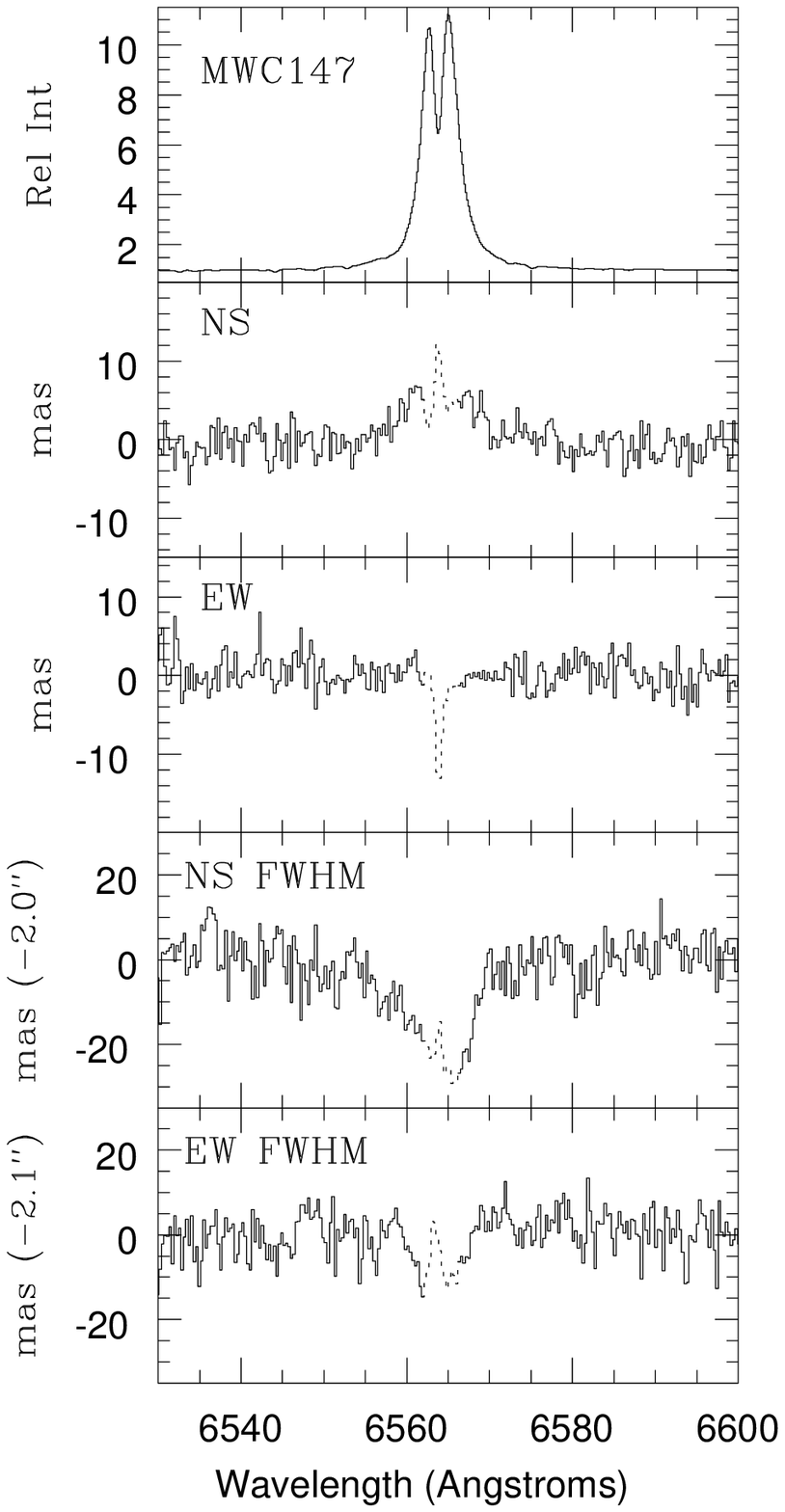,width=6.0cm}\\
\epsfig{file=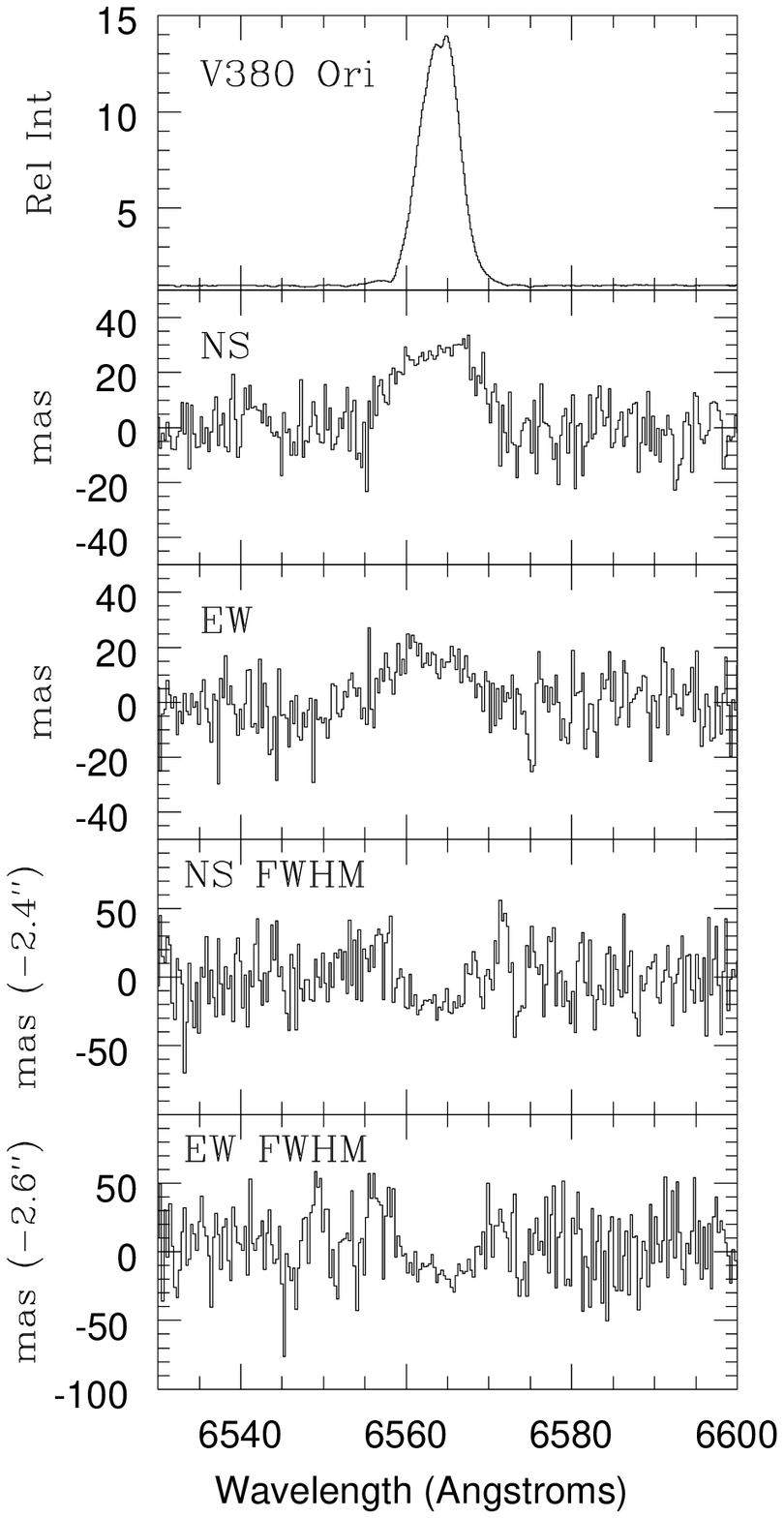,width=6.0cm}
\epsfig{file=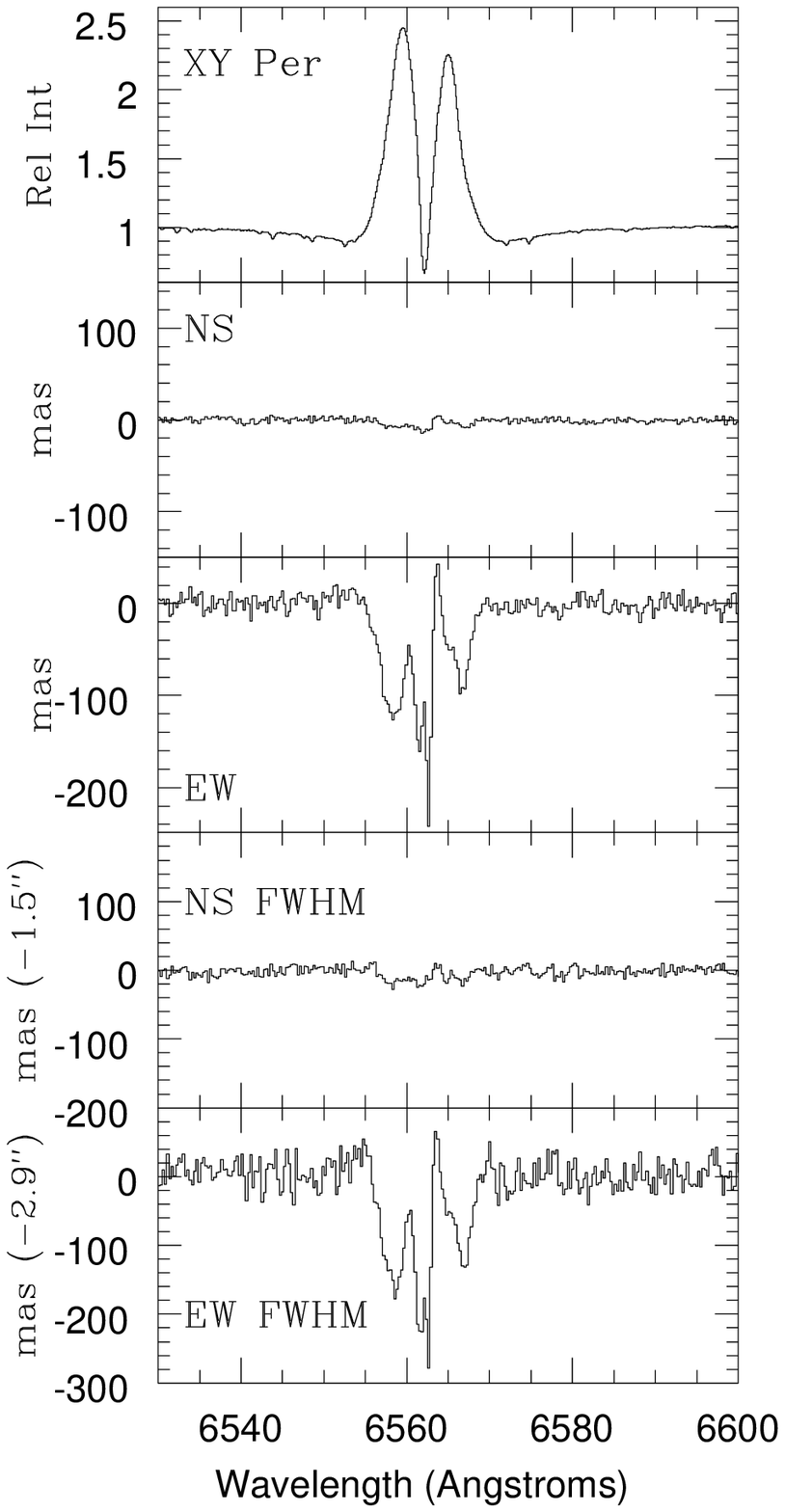,width=6.0cm}
\epsfig{file=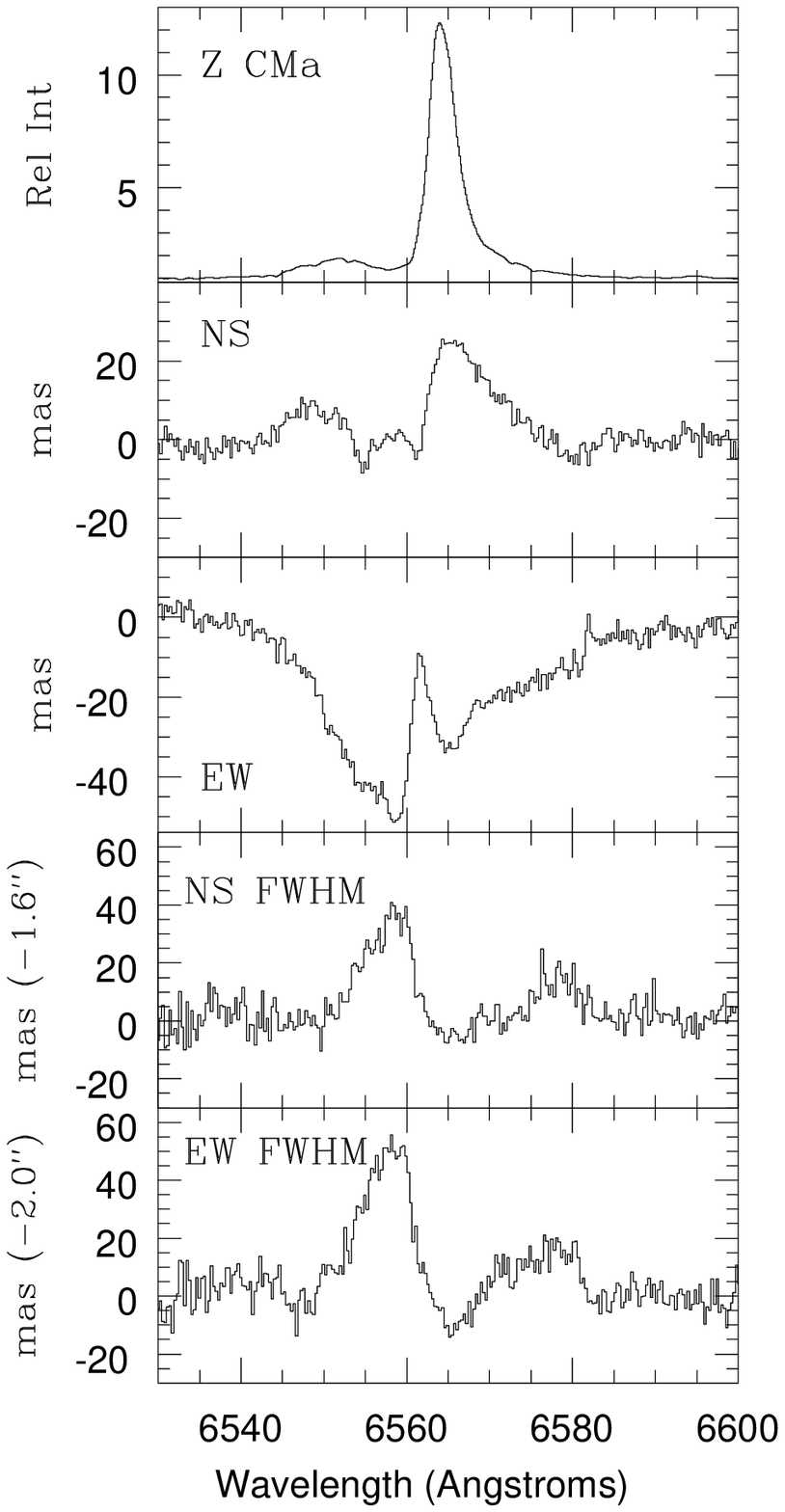,width=6.0cm}\\
\caption{As the previous figure, but now for the remaining known
binaries. Features that proved to be artifacts are indicated by a
dotted line (see text).
\label{binaryfig}}
\end{figure*}

Nine of the observed targets were previously known to be binaries with
a largest observed separation of 3.1 arcsec. Their data are presented
in Figures~\ref{GUCMa} and \ref{binaryfig} and their observed values are
listed in Table~\ref{binarytable}.  

HK Ori is extensively discussed in Baines et al. (2004) and is not
shown here.  V380 Ori was also observed by Bailey (1998), who found a
PA = 206.7 $\pm$ 1.1$^{\circ}$, and a positional displacement of 0.039
$\pm$ 0.003 arcsec to the North-East. Our results are consistent with
Bailey (1998) and are an important check on the reliability of our
data and the repeatability of the technique.  The PA listed in
Table~\ref{binarytable} is measured over the part of the spectrum that
shows an obvious binary signature. XY Per was observed on 2
consecutive nights. Although the H$\alpha$ profile varied on these two
days, we only show the first spectrum. The PA derived from both dates
is the same. Z CMa was also observed by Bailey (1998) and will be
discussed later.

In general, all of these objects show the spectro-astrometric
signatures of a binary system.  As outlined in the Appendix, the
multiple peaks in the position and FWHM spectra of XY Per can even be
explained by both components in the binary emitting H$\alpha$.  In
addition, for 7 of the binaries the position angles derived from the
spectro-astrometry agree very well with the existing data
(Table~\ref{binarytable}).  The object for which the PA was not
reproduced is one of two objects with by far the largest separations
(MWC~361 with a separation of 2.25 arcsec), much greater than the slit
width.  It has the smallest positional shift in our sample, and we
confirmed the reality of the signature.  It is only due to the
comparatively large seeing values that we detect its secondary at all.
MWC 147 also has a large separation, and care has thus to be taken in
interpreting its derived PA.  The binary has magnitude difference of
$\Delta R$ = 6.82 mag, but an unpublished PA (Corporon, 1998, PhD
thesis, cited by Millan-Gabet, Schloerb \& Traub 2001).  The fact that
this binary, which has a separation of 3.1 arcsec and an $R$ magnitude
difference of almost 7 mag, can still be detected reveals how sensitive
spectro-astrometry is at detecting binaries.

The empirical lesson learnt here is that in the case of wide binaries,
(conservatively defined here as having separations larger than half
the slit size, i.e. $> \sim$ 0.5 arcsec) we must be careful with any
interpretation from the calculated PA. Indeed, as noted by Porter et
al. (2004), when the intent is to study binary systems and their
secondaries, spectro-astrometric data has to be obtained with a very
wide slit. The trade-off is a significant loss of spectral resolution.

Another important point to note is the observed change in FWHM across
the H$\alpha$ line. In Table~\ref{binarytable} and
Fig.~\ref{binaryfig}, we find that the change is small for both small
(HK Ori, V380 Ori) and large (MWC 147, MWC 361) separations and
reaches a maximum in between. This can be understood in the following
way. The large separation binaries become resolved in our data and the
profile fitting routine will choose to fit the FWHM of only one
component, across both the continuum and the line. In the case of
small separations, the point spread functions of both components fall
on top of each other and therefore result in only a small change in
FWHM even if only one component is measured across the H$\alpha$ line.
Several variables come into play when interpreting the data and a
first attempt to quantify the effects is presented in the
Appendix. The empirical result that a significant drop in FWHM occurs
when the binary separation is comparable to the slit size is confirmed
by the simulations.

\begin{figure*}
\centering
\epsfig{file=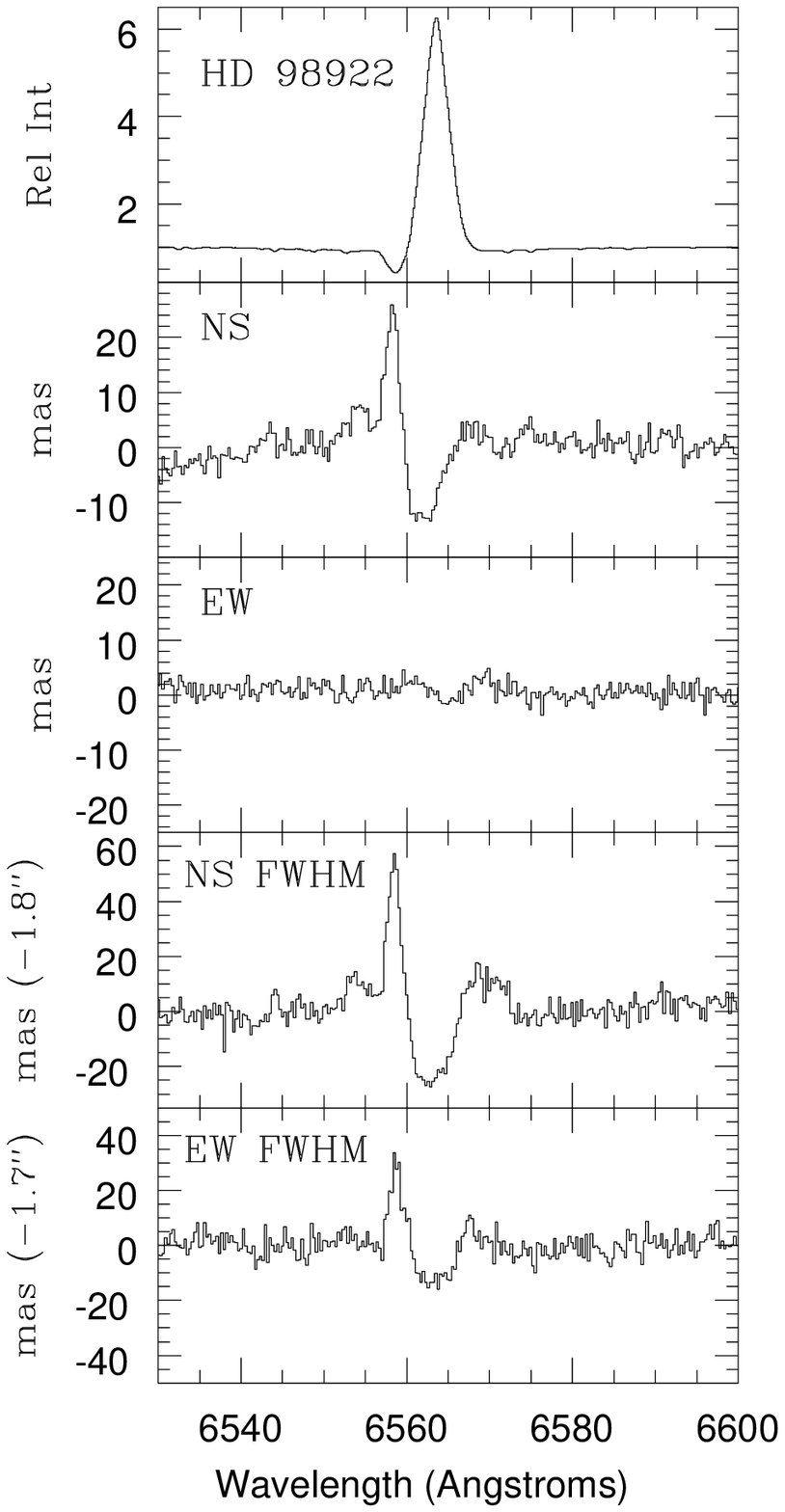,width=6.0cm}
\epsfig{file=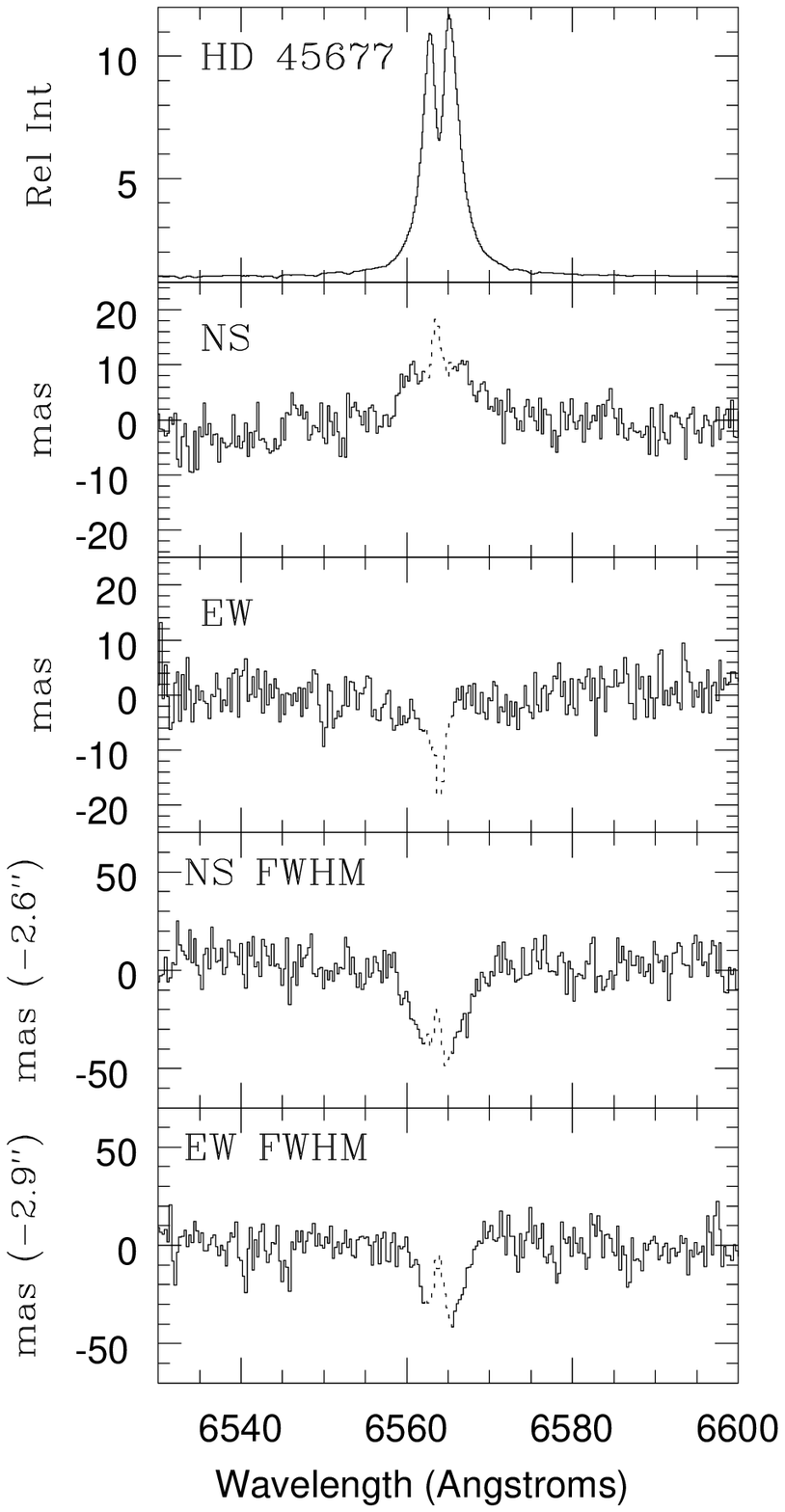,width=6.0cm}
\epsfig{file=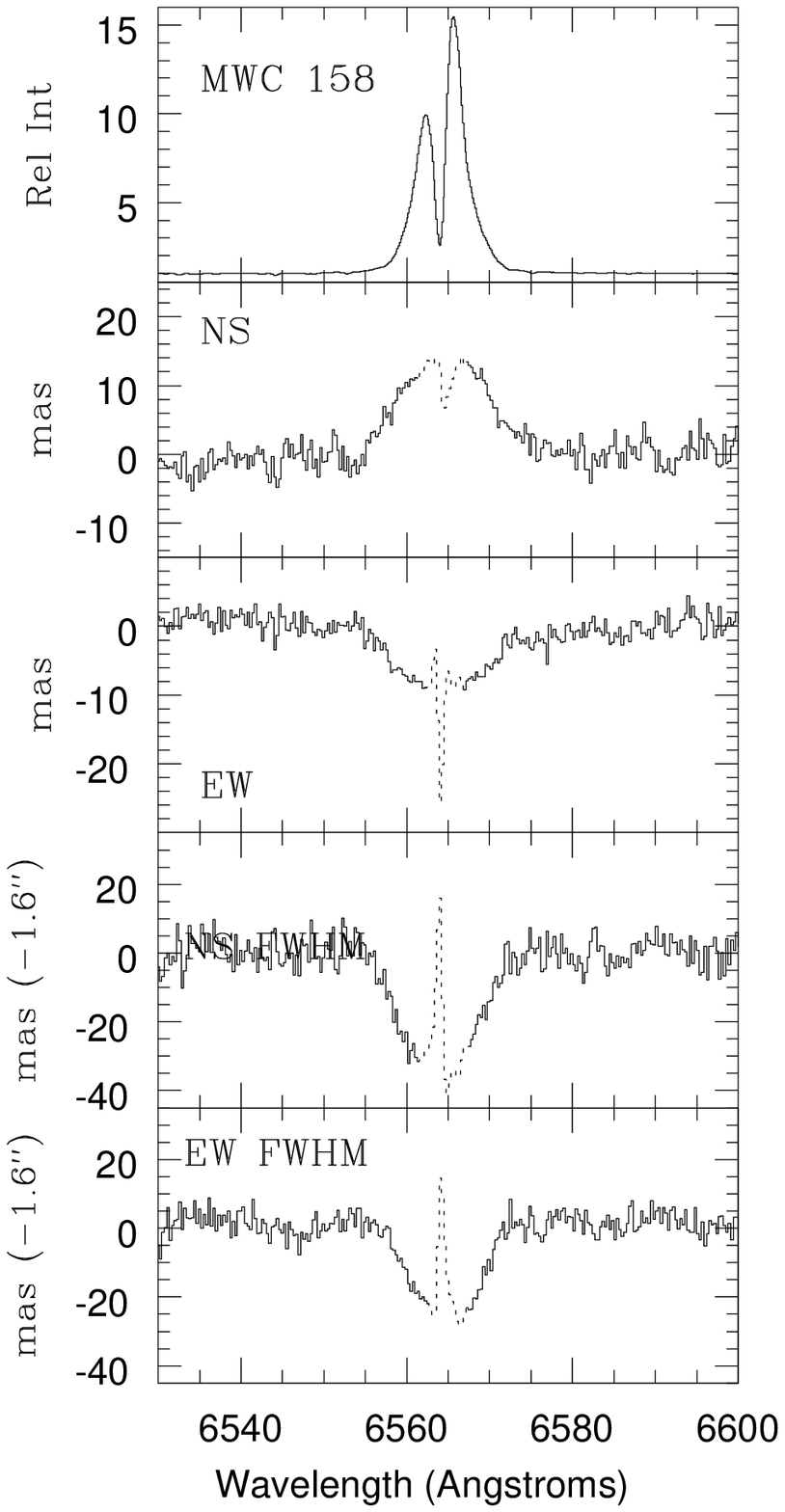,width=6.0cm}\\
\epsfig{file=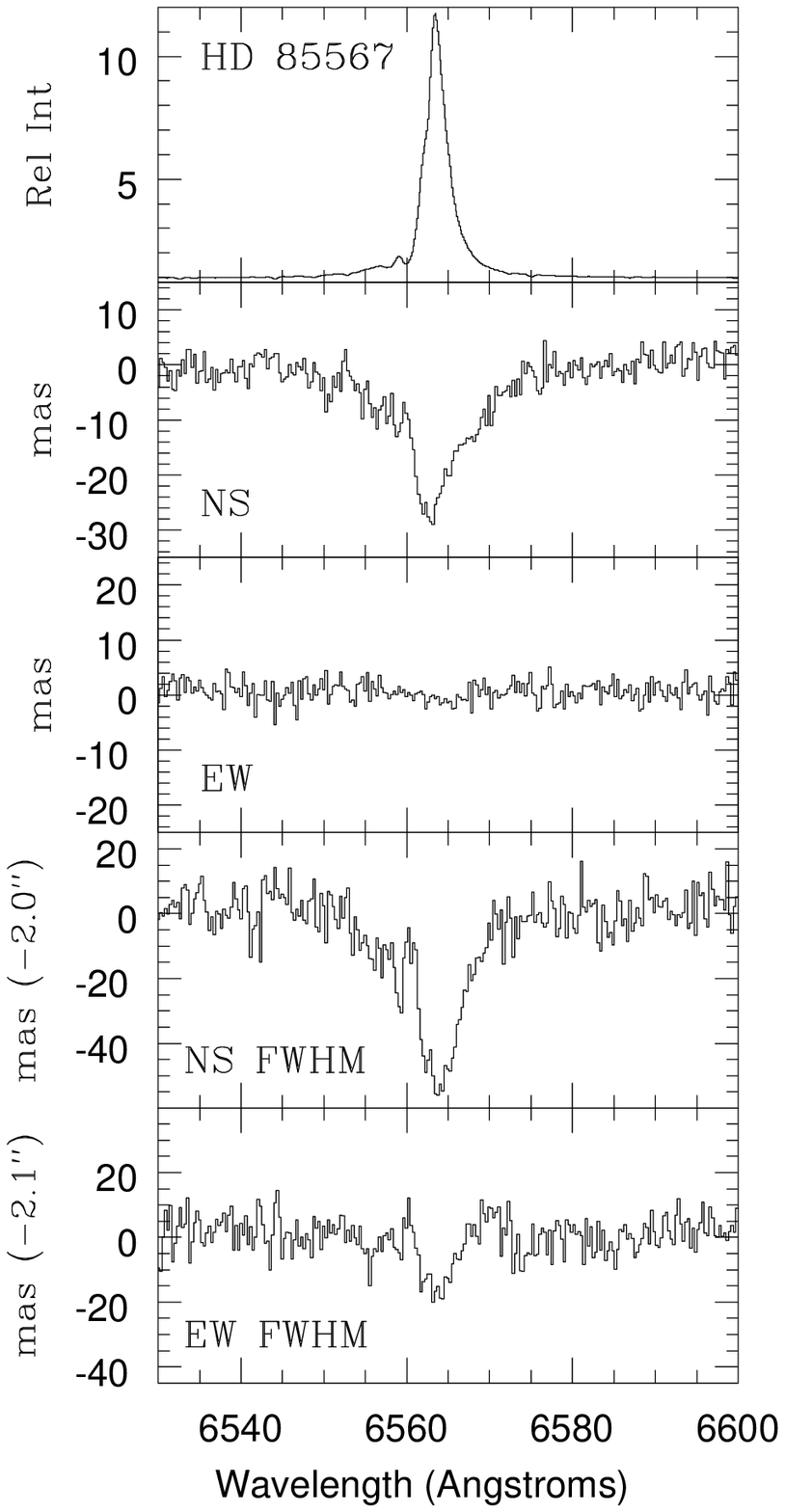,width=6.0cm}
\epsfig{file=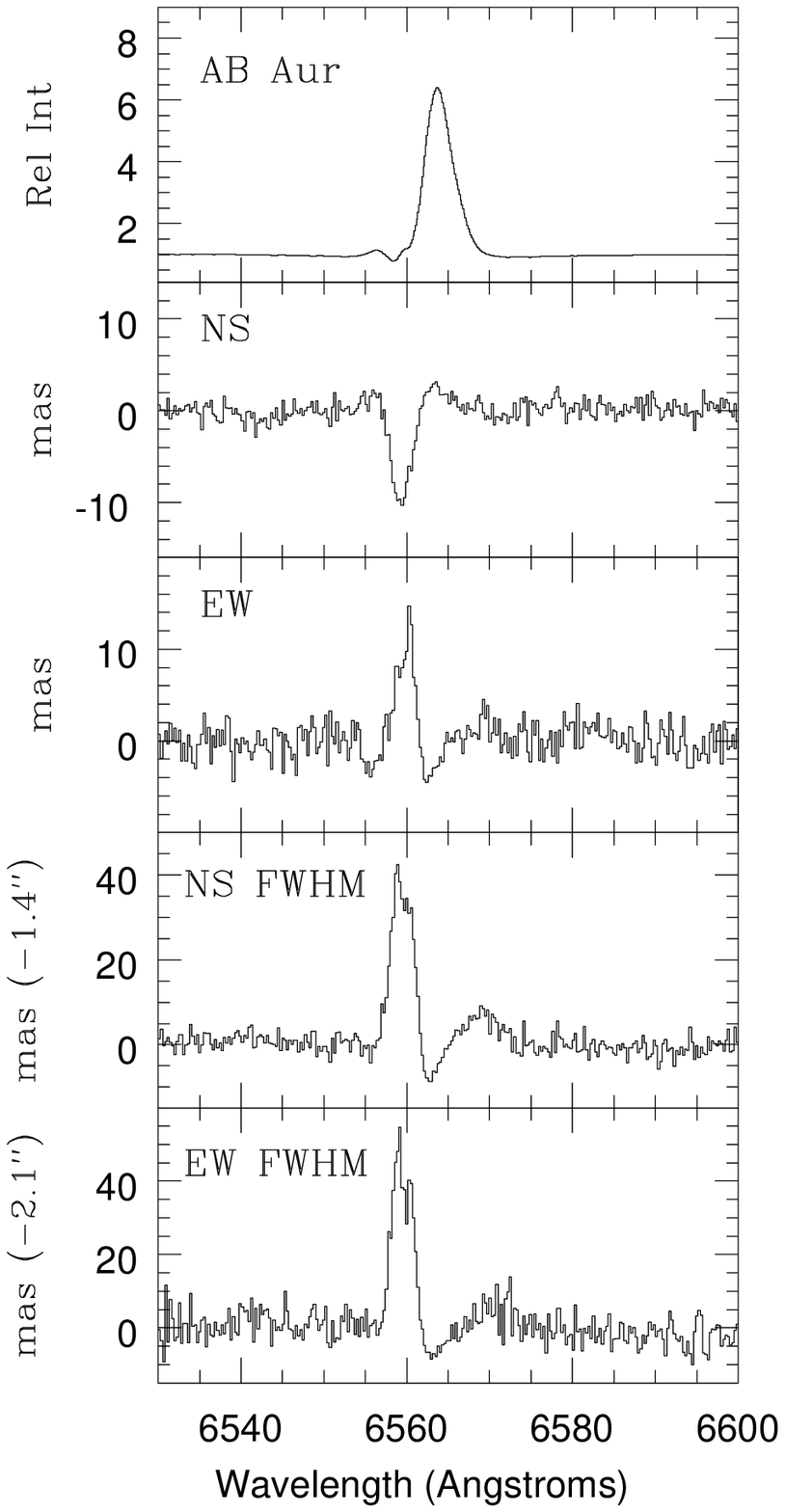,width=6.0cm}
\epsfig{file=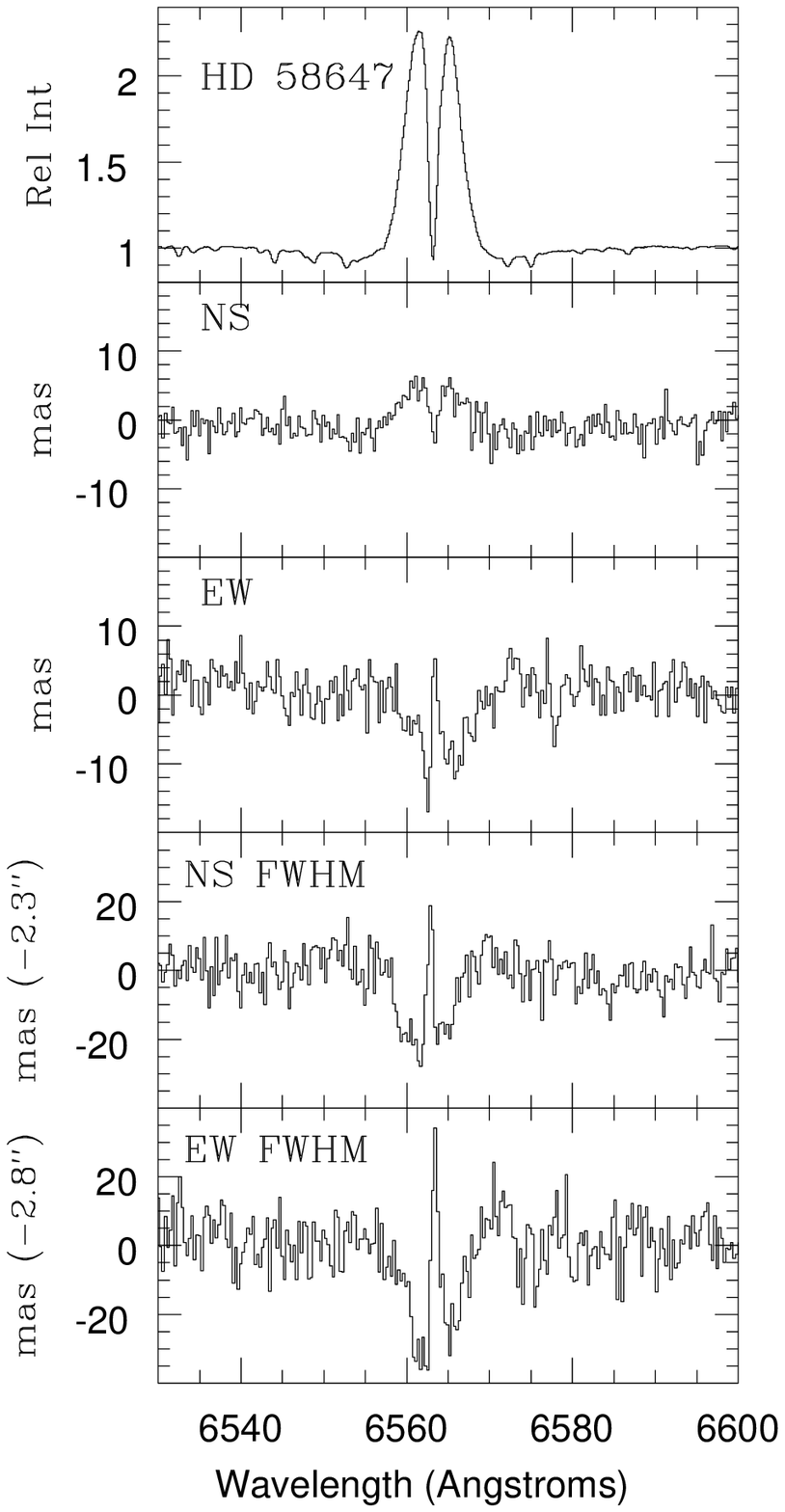,width=6.0cm}
\caption{As the previous figure, but now for the new binaries. 
\label{newbinary}}
\end{figure*}

\setcounter{figure}{2}
\begin{figure*}
\centering
\epsfig{file=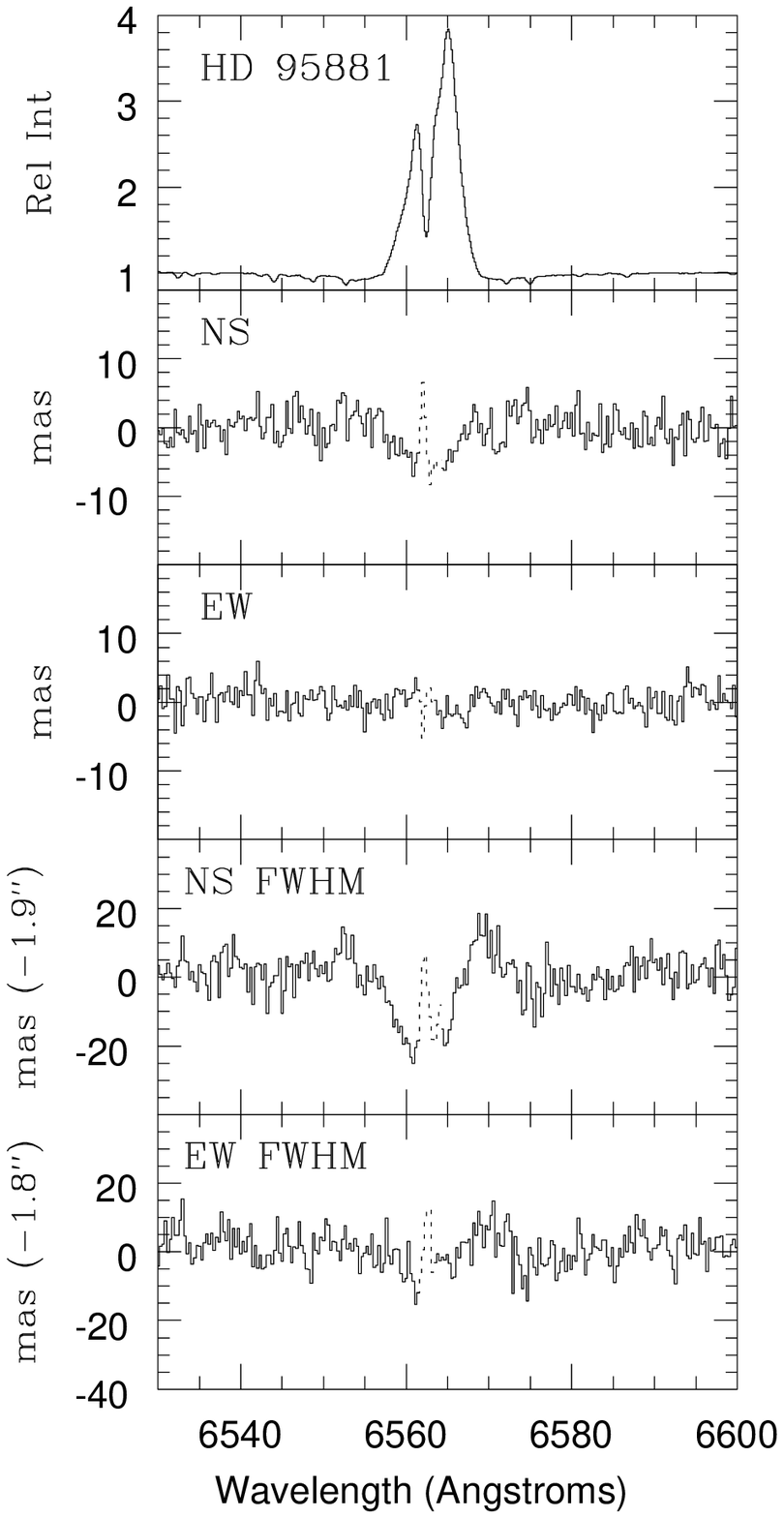,width=6.0cm}
\epsfig{file=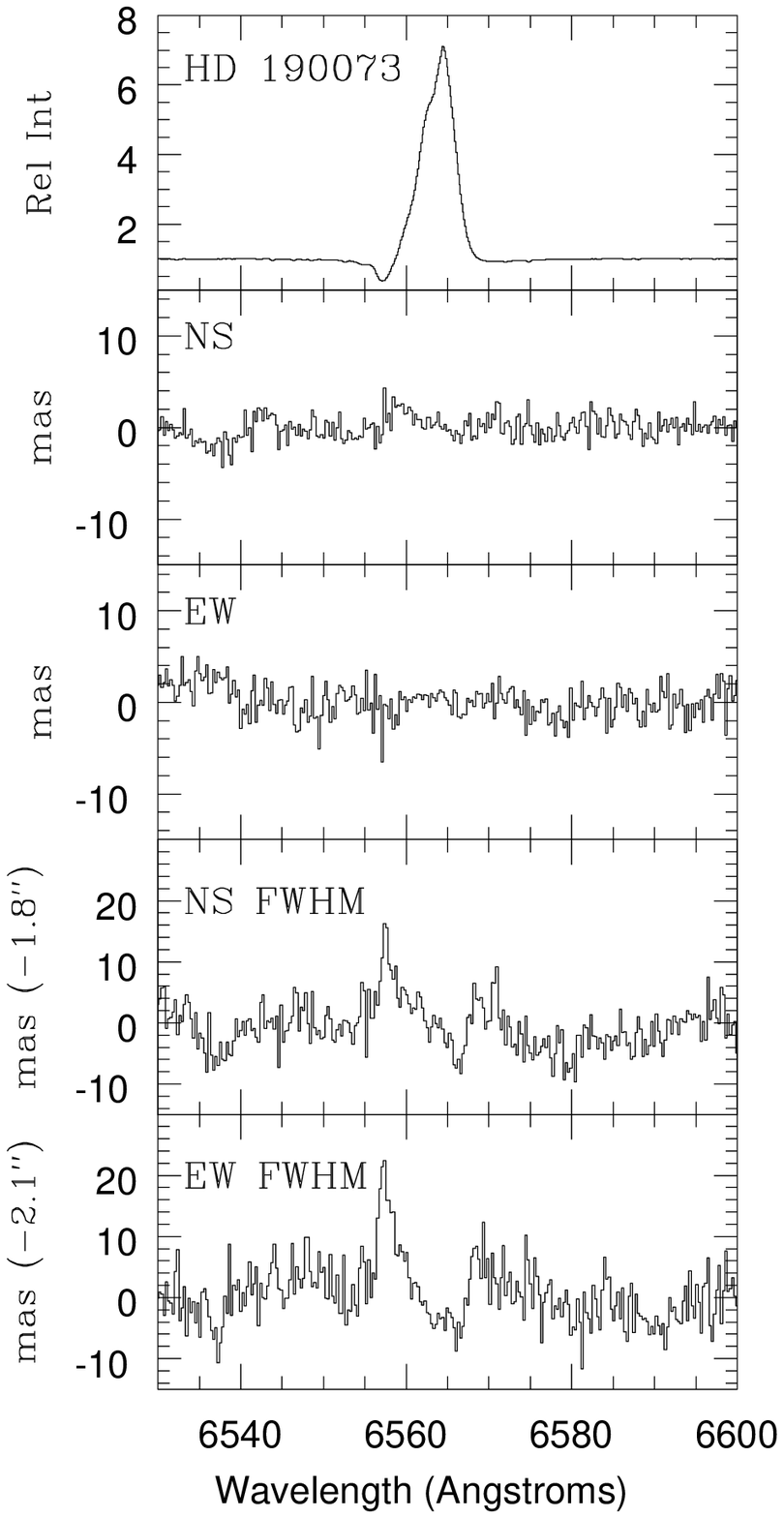,width=6.0cm}
\epsfig{file=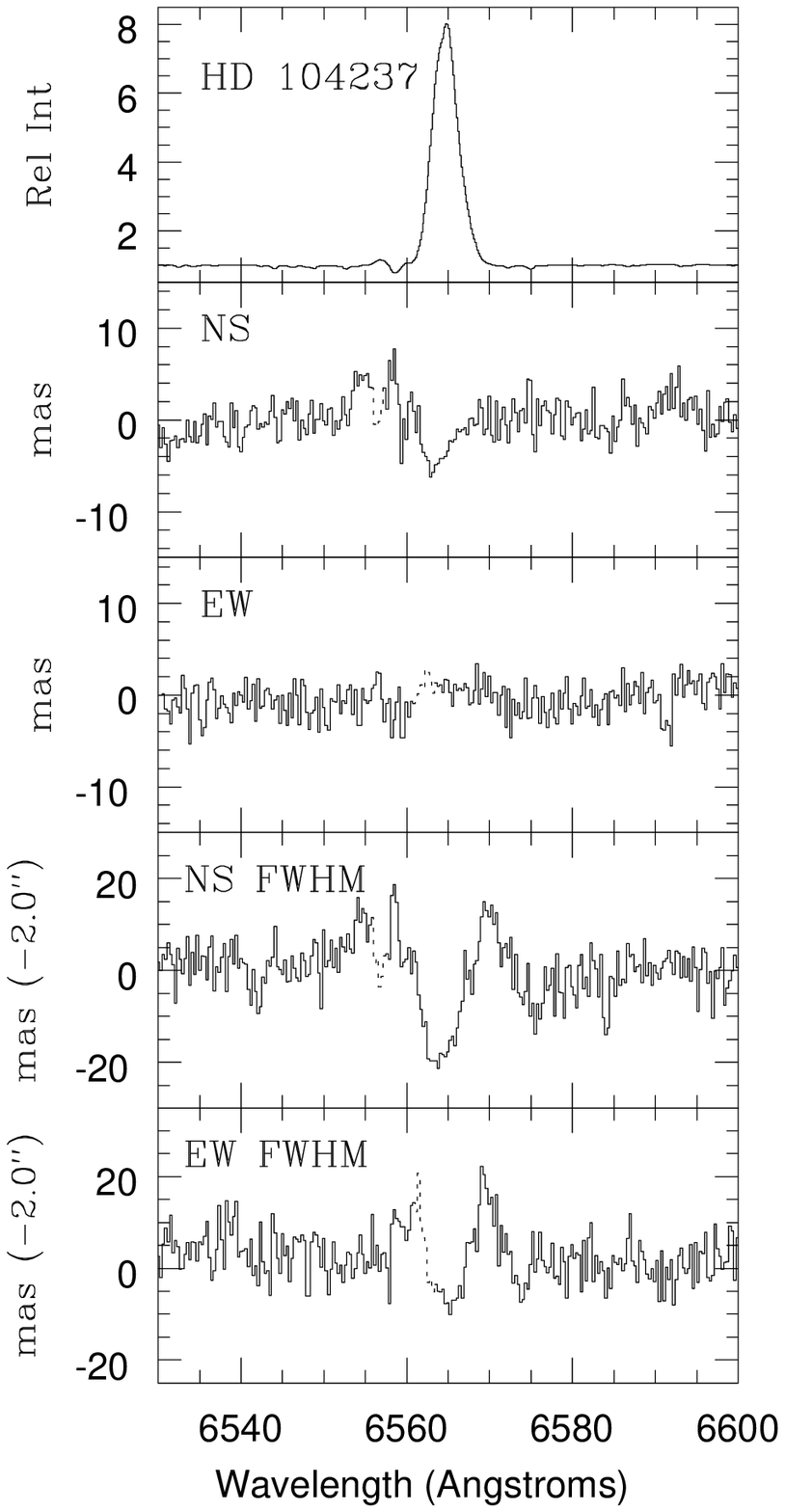,width=6.0cm}\\
\epsfig{file=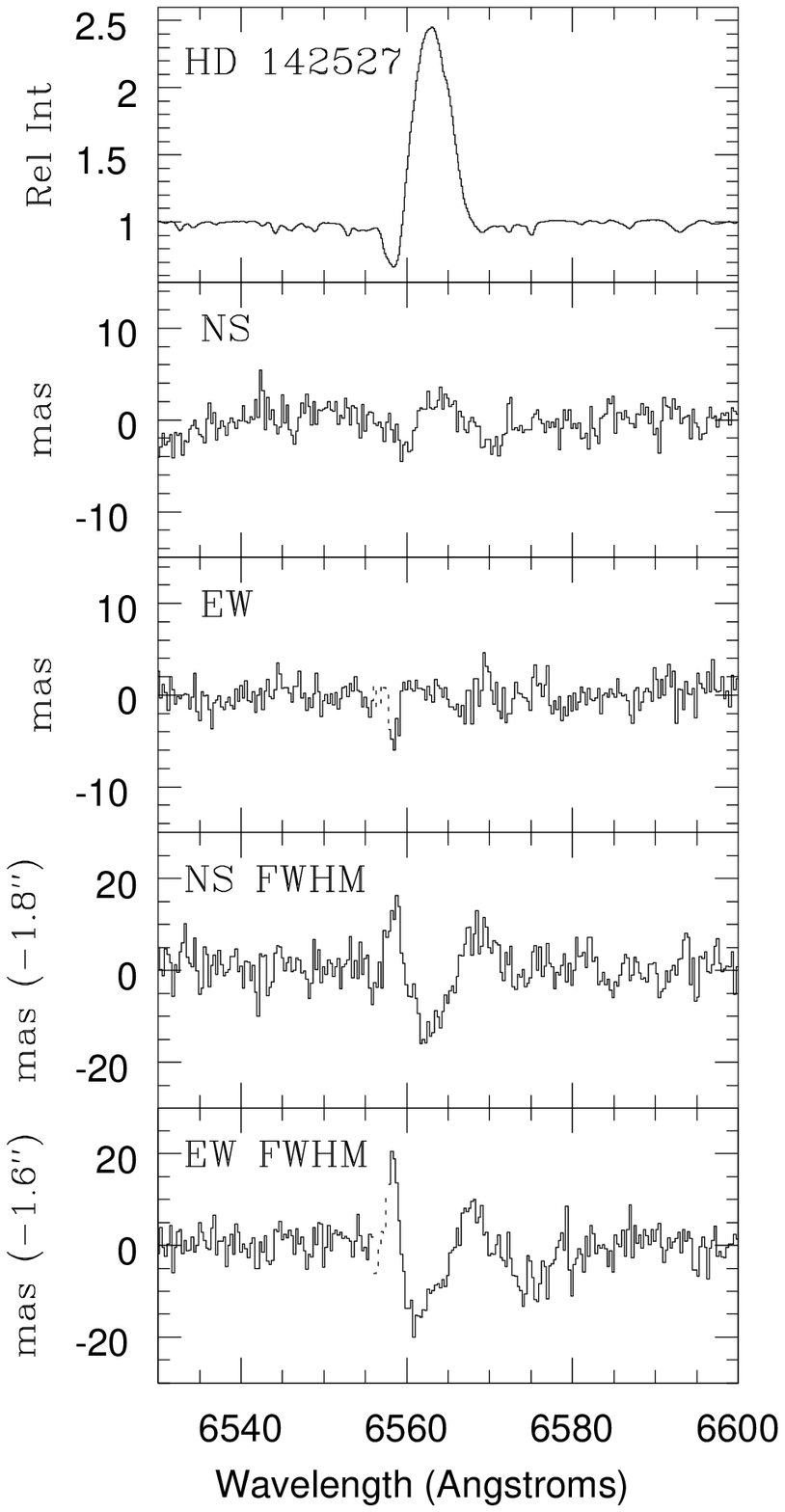,width=6.0cm}
\epsfig{file=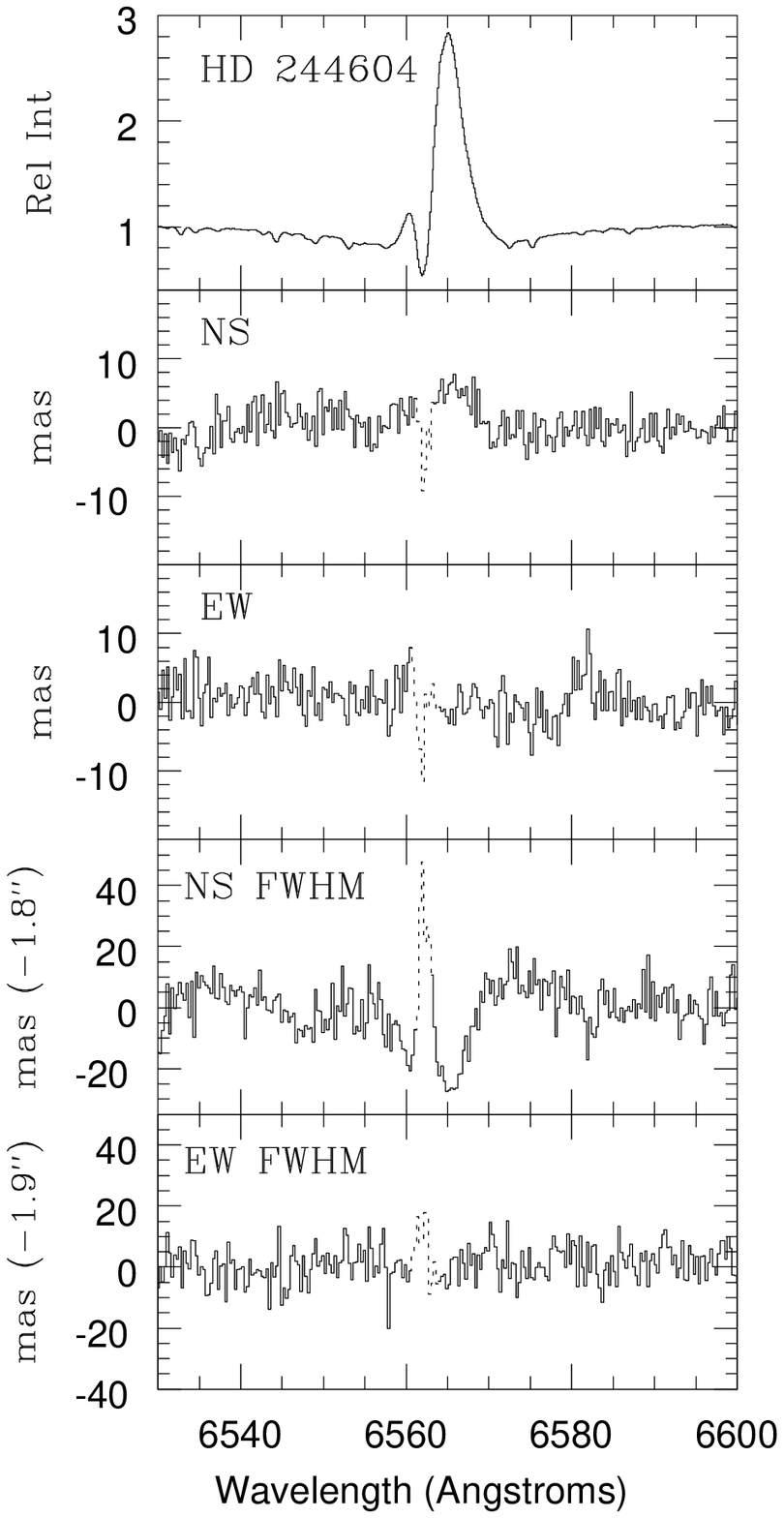,width=6.0cm}
\caption{Continued - Possible binaries \label{newbinary2}}
\end{figure*}

\subsubsection{New binary systems}\label{NewBins}

Within the dataset we find six stars which display a signature in the
positional and FWHM spectra consistent with binarity of the objects
(AB Aur, HD 45677, HD 58647, HD 85567, HD 98922, and MWC 158).
According to our knowledge these are all new detections. Their data
are presented in Fig.~\ref{newbinary} and derived quantities are
listed in Table~\ref{binarytable}. As all of the objects show obvious
changes in the FWHM spectrum it follows from our finding outlined
above and reinforced later, that the binary separations in all cases
are greater than half the slit width ($>$ 0.5 arcsec). We can thus
conclude that the companions of these wide systems are most likely at
a separation of 0.5 to about 3.0 arcsec, the separation where the
change in FWHM starts to decrease.

Both AB Aur and MWC 158 were observed during the two observing runs at
different telescopes, and provide an internal consistency check. AB
Aur qualitatively displays similar features on both occasions. This is
important as it yields confidence in the reality of the detected
features. However, the displacements are larger in January than in
September, while the derived position angle is rather different (see
Table~\ref{binarytable}). This is consistent with the idea that the
binary has a larger separation than the slit width.  Indeed, the seeing
was slightly better in September, the light from the secondary
therefore contributed less to the total spectrum. The result is that
smaller excursions in the spectro-astrometric data occur, as observed.
It also implies that the derived position angles from such binaries
are highly uncertain.

An even more extreme effect occurred for MWC 158, whose binary
signature in September all but disappeared.  If it were not for the
agreement of the AB Aur data and the other tests outlined earlier, it
could be argued that the method is flawed. However, it is probable
that in the better seeing the light from the secondary did not
contribute enough to the total light to influence the
spectro-astrometric data. In small seeing this effect is even stronger
if the primary is not centered exactly in the centre of the slit.

We also wish to draw attention to HD 98922.  This star is a known
binary, but at a large separation of 7.8 arcsec, and a PA of
343$^{\circ}$ (Dommanget \& Nys, 1994).  If we assume that the primary
is well centred, then its companion is located almost 2 arcsec outside
the slit in the NS direction and about 7 arcsec in EW. As the seeing
during the observations was 1.7 arcsec it is not impossible that we
could just be detecting this binary in the North-South direction but
it would be very difficult to detect the secondary in the
East-West direction, especially in the FWHM spectrum. We therefore
include this object in the list of new detections, rather than list it
as a recovery of a known object.

\subsubsection{Possible binary detections}\label{Possbins}

Five stars within our sample show spectro-astrometric signatures that
suggest a binary detection. These stars do not, or hardly, display any
feature in the positional spectrum, but do show FWHM displacements
consistent with a binary (HD 95881, HD 104237, HD 142527, HD 244604
and HD 190073). As the evidence for a binary is only visible in the
FWHM spectrum, we prefer to be conservative and list these detections
as ``possible'' binaries.  The fact that the FWHM data display a
larger excursion than the position spectra, indicates that the
separation of the possible binaries is larger than half the
slit-width. No position angles could be derived for these objects as
the excursions in the positional spectra can not be measured.

\subsection{Objects with evidence for outflow}

\subsubsection{Z CMa}
\begin{figure}
\centering
\epsfig{file=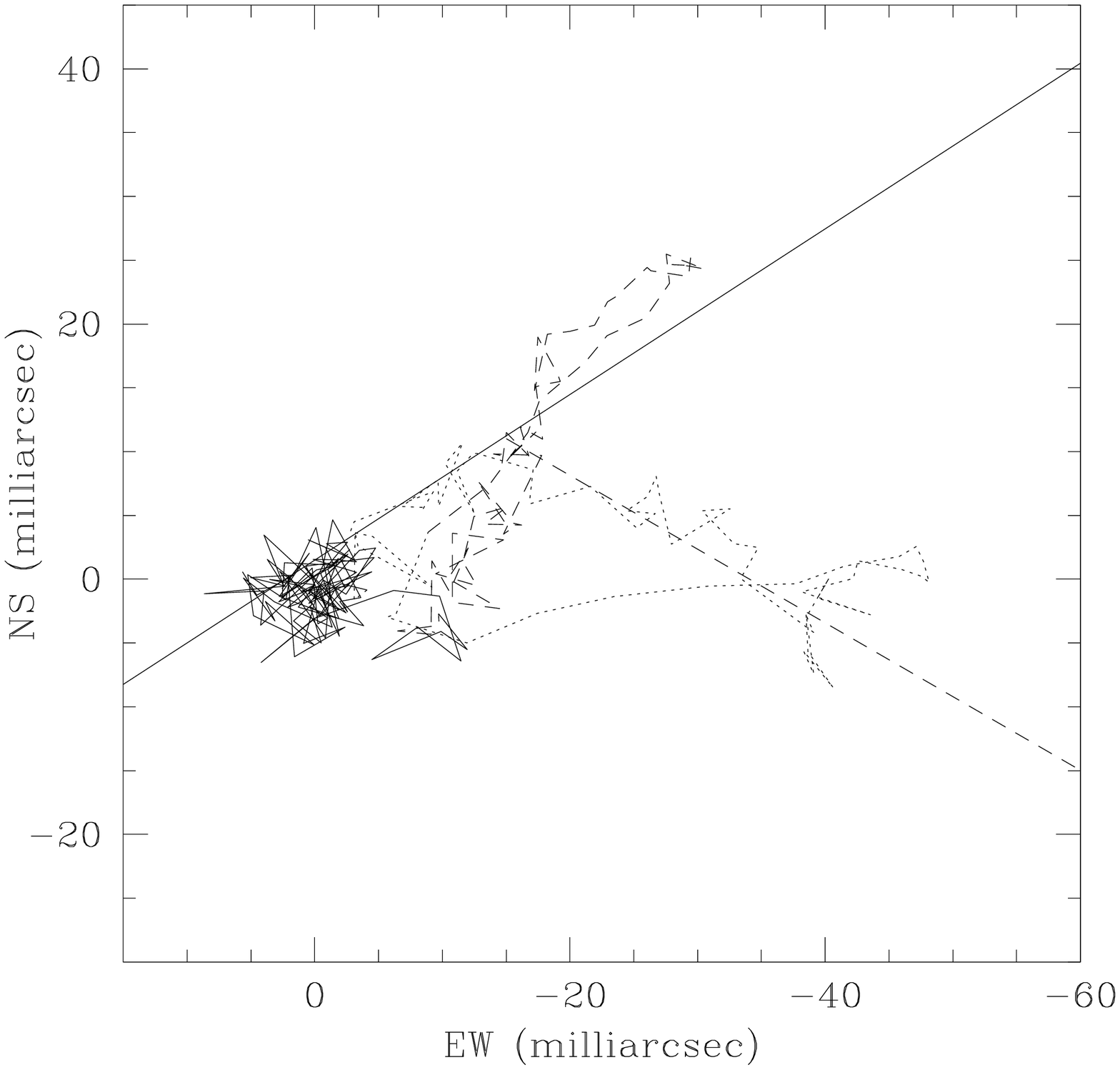,width=8.0cm}
\caption{The spectro-astrometric $XY$ plot of Z CMa, centred on
H$\alpha$. The solid line shows the PA of the binary (122.5$^{\circ}$),
the dashed line shows the PA of the extended outflow (240$^{\circ}$).
Red and blueshifted components across the H$\alpha$ profile are
plotted as a dashed line and a dotted line respectively, to illustrate
which of these are a result of the outflow, and which are a result of
the binary. The blueshifted outflow is traced by
the blueshifted positional displacements, except the far
blue wing, which traces the binary. North is up and East to the left.
\label{ZCMa}}
\end{figure}

ZCMa was also observed by Bailey (1998) who found that the
spectro-astrometry of the linewings traces the (known) binary while the
jet emission in the line centre traces a well known outflow (see also
Garcia Thi\'ebaut \& Bacon, 1999). Although the emission has changed
with respect to his spectrum, the positional displacements we observe
are similar to those found by Bailey (1998).

The data are represented in a so-called XY-plot in Fig.~\ref{ZCMa}.
This graph plots the excursions in the EW and NS direction against
each other.  The red-shifted emission follows the binary direction,
and the derived PA corresponds to the binary PA of 120$^{\circ}$
(Koresko et al. 1991; Leinert et al 1997).  The blueshifted part of
the line is dominated by jet emission and causes the positional
displacements to move off the binary line in the direction of the jet
at a PA of 240$^{\circ}$ (Poetzel, Mundt \& Ray 1989).  The FWHM
spectrum (Fig. 2) also indicates the presence of a binary; the
redshifted displacements are consistent with the features being due to
a binary as the FWHM decreases over the emission, and increases over
the P Cygni absorption. There is an increase in the FWHM across the
region dominated by the jet.  From the direction of the binary
displacement we can infer that the brightest infrared star also
dominates the H$\alpha$ emission.

\subsubsection{HD 87643}

HD 87643 stands out in the complexity of its spectro-astrometry.
Oudmaijer et al (1998) conducted an extensive study of the star by
optical spectroscopy, spectropolarimetry and imaging.  HD 87643 shows
the B[e] phenomenon and also some characteristics typical for pre-main
sequence stars. The spectropolarimetry revealed the presence of a
rotating, expanding small scale ionized disc.

\begin{figure}
\centering
\epsfig{file=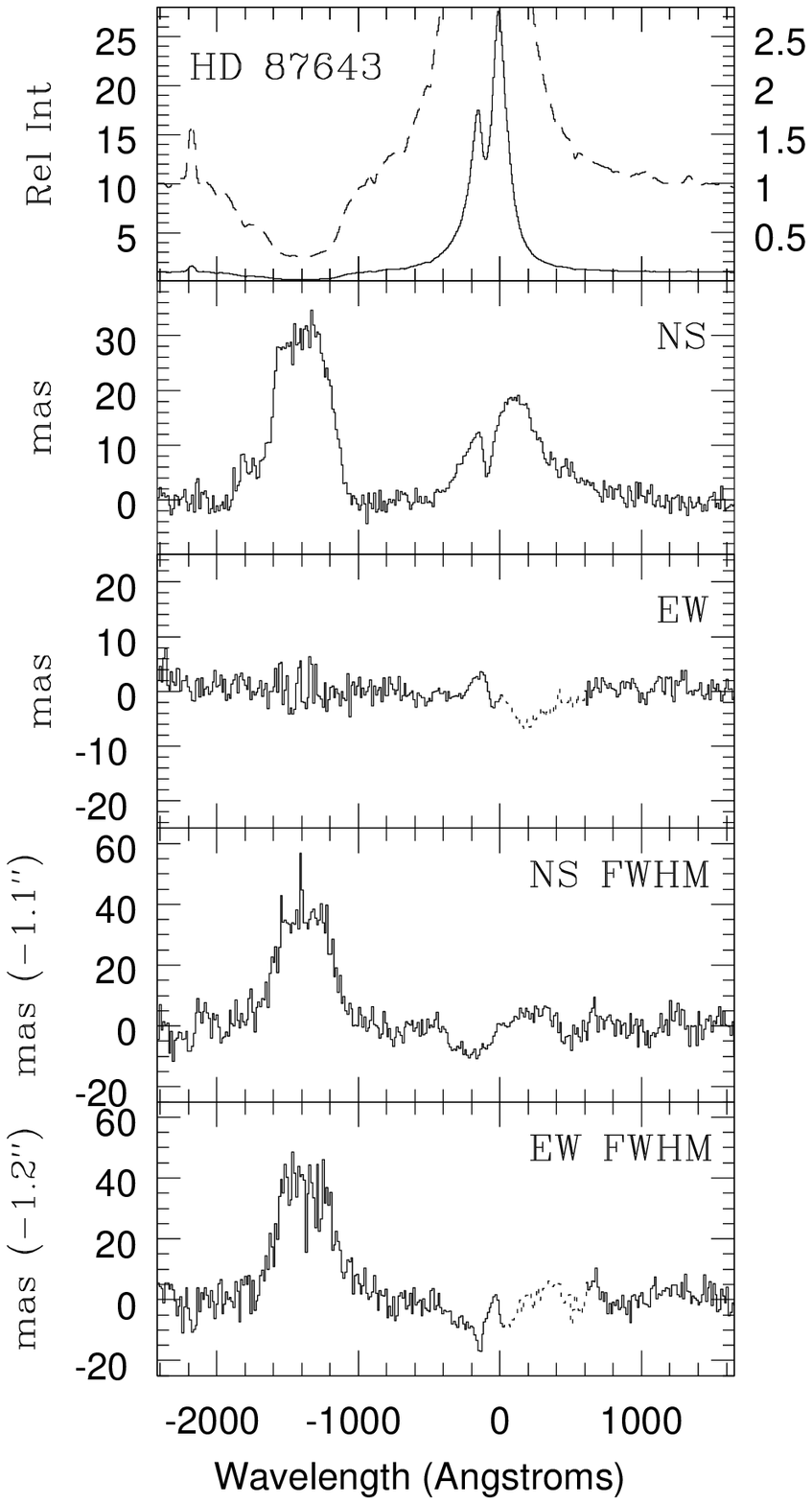,width=6.0cm}
\caption{The spectro-astrometric plots of HD 87643, centred on
H$\alpha$. The dashed line in the top panel is a magnification of the
spectrum, to illustrate the P Cygni absorption.
\label{HD87643}}
\end{figure}

\begin{figure}
\centering
\epsfig{file=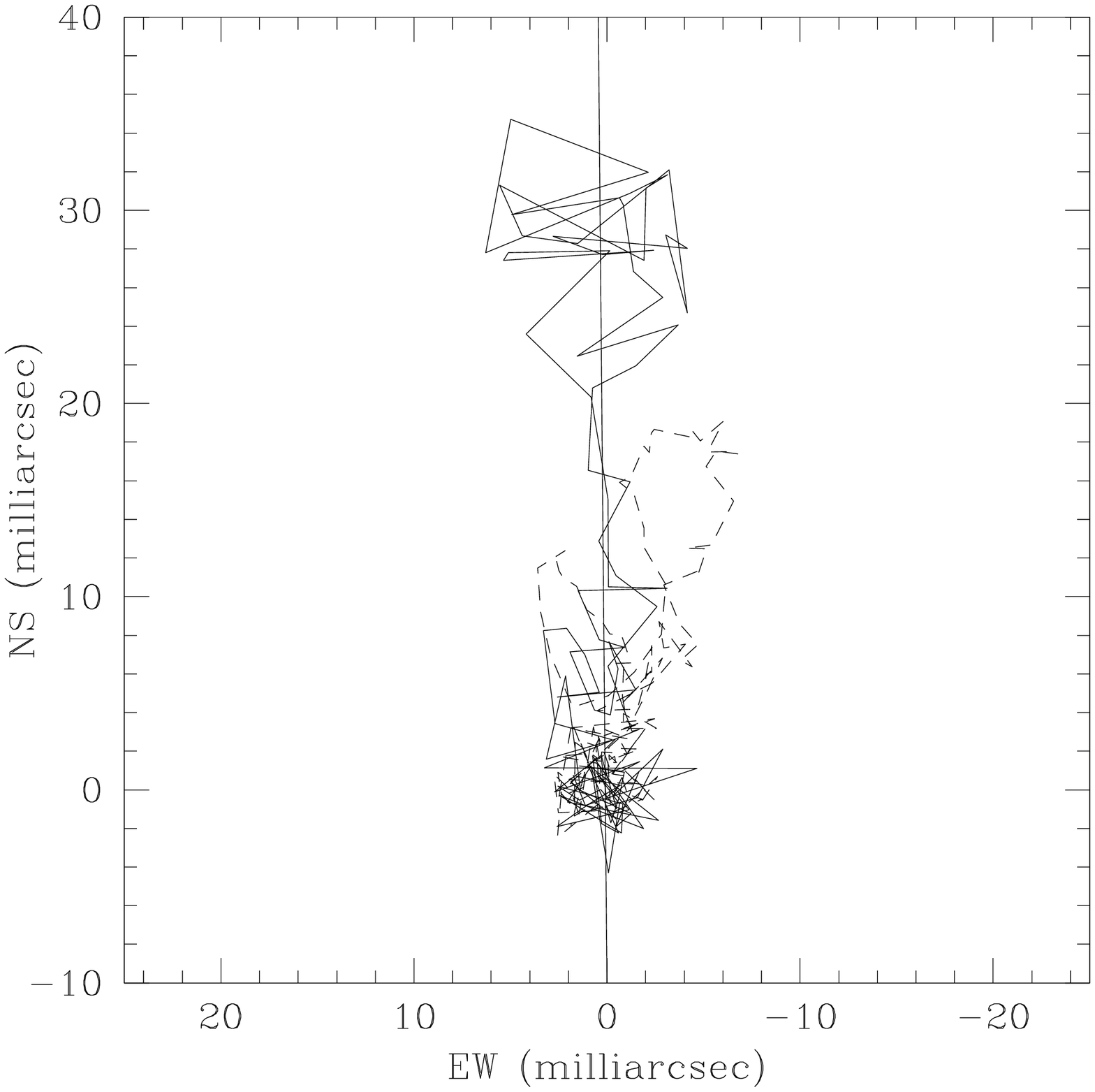,width=8.0cm}
\caption{The XY plot of HD 87643 centred on H$\alpha$, between 6520 to
6590 $\rm \AA$. The displacement across the P Cygni absorption is
shown as a solid line, and across the emission as a dashed line.
\label{HD87643XY}}
\end{figure}

Figure \ref{HD87643} displays the spectro-astrometric plots of HD
87643. A northwards positional displacement  across the P Cygni
absorption is accompanied by a large increase in both the NS and EW FWHM
spectra.  There is no corresponding feature in the EW position
spectrum however.  Across the H$\alpha$ emission line itself, a
significant positional displacement northwards is present, but little
change in the FWHM spectra.  The XY plot is presented in
Fig.~\ref{HD87643XY} from 6520 to 6590 $\rm \AA$. This reveals that
both the displacement across the P Cygni absorption (solid line) and
emission (dashed) fall along the same PA of 1.7 $\pm$ 0.9$^{\circ}$.

The simplest scenario, a binary system, fails to explain the data. The
NS position spectrum shows a displacement across the P Cygni
absorption, as already described for some binaries earlier.  However,
in contrast to those cases, the positional displacement across the
H$\alpha$ emission is in the same direction as that of the
absorption. This would imply that the secondary dominates the
H$\alpha$ emission, but the primary would  have the P Cygni
profile. Therefore, both primary and secondary would be emitting
H$\alpha$.  Although contrived, this situation is not
impossible. However, in the EW direction the argument for a binary
system breaks down. No positional displacement is observed across the
P Cygni absorption, but the EW FWHM still increases across the P Cygni
absorption - contrary to what is expected for a binary, and contrary
even to what is observed in the NS direction.

A second scenario, is that the H$\alpha$
emission is due to an outflow. An increase in both the NS and EW FWHM
across the P Cygni absorption could be expected from an extended
outflow. The absence of a positional displacement across a feature, in
this case the P Cygni absorption in the EW,  simply indicates that
the outflow is symmetric and extended in that direction.  At the same
velocities within the P Cygni absorption, a displacement and an
increase in FWHM is observed towards the North, indicating that this
outflow is asymmetric in the NS direction. The H$\alpha$ emission also
shows a displacement towards the North, although not as large as in
the absorption. No increase in the FWHM is seen across the H$\alpha$
emission, and presumably this structure is closer to the star.

From these spectro-astrometric displacements we suggest that we are
observing two different but aligned velocity outflows.  Across the P
Cygni absorption we detect a high velocity outflow, symmetric in the
EW direction and asymmetric towards north. Closer to the star is an
extended H$\alpha$ emission region, also with an extra component
towards the north.

Such high velocity positional displacements could be produced by a bow
shock type structure towards the north. The low velocity positional
displacements may be from the same outflow as the high velocity
displacement, but closer to the star.  The polarization of HD 87643
shows a disc-like structure surrounding the star with polarization
angle of $\sim$ 20$^{\circ}$ and is probably viewed at a high
inclination angle (Oudmaijer et al., 1998).  Our results imply a high
and low velocity outflow predominantly located towards the north (PA =
0$^{\circ}$). This is closer to the PA of a polar outflow ($\sim$
20$^{\circ}$) than the disc when assuming the disc lies perpendicular
to the polarization.  Finally, we note that the spectro-astrometric
displacements of HD 87643 stand out as different to all our other
Herbig Ae/Be stars, and hence HD 87643 may indeed belong to the B[e]
supergiant group.

\section{Discussion}\label{Diss3}

We have observed 28 Herbig Ae/Be stars and 3 F-type pre-main sequence
stars using spectro-astrometry.  The data prove both from an empirical
point of view and from simulations described in the Appendix very
powerful in discovering binary objects. The method exploits the fact
that the components have different spectra and is used here across the
H$\alpha$ emission line.  After a discussion on spectro-astrometry
itself, we will investigate the implications of the large binary
fraction of these intermediate mass pre-main sequence objects for
their formation, and investigate the alignment of the binary systems
with the circumprimary discs.

\subsection{Spectro-astrometry as a tool to investigate binary stars}

As this paper presents the first comprehensive study into binarity
using spectro-astrometry, we briefly discuss the possibilities and
limitations of the method here.  

\subsubsection{On the capabilities of spectro-astrometry}

Checks on known binary systems that were present in our sample - which
were all retrieved - and the repeat observations of objects at
different telescopes indicate that the method is robust.  The smallest
separation binary that was observed has a separation of 0.1
arcsec. The technique can cope with a large flux difference, both
simulations and observations (MWC 147) indicate that secondaries which
are up to 6-7 magnitudes fainter can still be detected.  The position
angles of the binary systems can be reliably reproduced for
separations smaller than the slit-width.  For larger separations, only
a fraction of the light from the secondary will fall within the
slit. This may result in a smaller positional displacement than
expected and an incorrect binary PA.  We should stress that the
position of the photo-centre only provides a lower limit to the true
separation. This is because it measures the position of one star
across the emission line, but a weighted average of the two stars in
the continuum. The magnitude of the displacement depends on the
seeing, separation and, crucially, flux-difference of the components.

The FWHM spectrum, which has never before been discussed in the
literature, provides  important  additional  information. It
gives diagnostic clues about the nature of the position spectra. The
stars with P Cygni profiles provided an illustrative example of the
different types of behaviour that can occur when observing absorption
or emission. The FWHM decreases over the emission, as only one object
is visible. In contrast, the FWHM increases over the P Cygni
absorption as the fainter secondary becomes more dominant in the
spatial profile.  In addition, the FWHM information also provides an
important indication of the binary separation. Binaries with
separations approximately larger than the slit width have FWHM
displacements as large as, or larger than, the position offset.

To learn more about the capabilities of the method, simulations were
performed and a selection of the results is presented in the
Appendix. The simulations reinforce the empirical results. The lower
limit to the binary separation is found to be smaller than the
smallest separation binary discussed here, and is of order $\sim$0.010
arcsec, with a dependency on the observing conditions and binary
parameters. Spectro-astrometry can detect secondaries up to 6
magnitudes fainter than the primary even for seeing values of $\sim$
1.5 arcsec.

As the separation is an unknown, it is virtually impossible to know
the flux losses induced by the limited slit width (see also Porter et
al. 2004). It is therefore not feasible to retrieve separations from
spectro-astrometric data without dedicated modelling while a large
slit width is required to ensure all light from the secondary is
captured.  For the same reason it is not possible to derive the
spectrum of the secondary for wide binaries with confidence. This
potentially important information for the understanding of binary
formation needs dedicated observations as discussed by Porter et
al. (2004).

\subsubsection{Detection of binary systems}

Among the 28 Herbig Ae/Be stars observed, 15 binaries (6 of them new
discoveries) have been detected. A further 4 Herbig Ae/Be and 1 F type
object are considered possible detections of a binary. 
The spectro-astrometry is able to reveal the presence of a binary only
if the respective spectra are different. In the present study we take
advantage of the fact that one of the components dominates the
H$\alpha$ emission. 

Before we begin interpreting the data, let us first consider selection
effects. The first and foremost question to ask is whether there is
any relationship between the strength of the H$\alpha$ emission and
the detection of a binary. By its very nature, the method is less
efficient for those objects with small spectral differences and hence
those objects with weak H$\alpha$ emission. Indeed, the 3 objects with
positive H$\alpha$ equivalent widths, indicating very weak emission if
at all, SV~Cep (A-type), BH Cep (F-type), and 51 Oph (B-type) are not
found to be a binary. For these objects it is reasonable to assume
that the method would not have revealed a binary even if a companion
were present. For the remainder of the sample, it would appear that
the method is not biased against faint H$\alpha$ emitters. The next
weakest H$\alpha$ emitter is MWC 166. Although its emission is only
40\% above the continuum (Fig.~\ref{binaryfig}), the binary is a clean
detection. The strongest H$\alpha$ emitter (MWC 137) turns out to be a
non-detection, while there is no obvious difference between the
strength of detections and non-detections. We therefore conclude that
the statistics are not significantly affected by the strength of the
emission line.

Another issue is the signal-to-noise ratio of the data. The higher SNR
spectra could in principle be more efficient at detecting signatures.
The two objects with the worst SNR in the position spectra are HK Ori
and V380 Ori but are both binary detections, while generally, both
detections and non-detections have on average comparable SNR. As there
is a narrow range in the rms variations in the photo-centre of the
data (1-8 milli-arcseconds, with most below 3 mas, see
Table~\ref{binarytable}) to first order the quality of the data has no
obvious effect on the results.

\subsection{On the binarity of Herbig Ae/Be stars}

The observations return a Herbig Ae/Be binary frequency of 54 $\pm$
11\% (68\% confidence interval). The fraction rises to 68 $\pm$ 11\%
if we include the possible detections.  Recent studies of Herbig Ae/Be
stars have have observed binary frequencies of 25 - 40\% (using IR
imaging: Bouvier \& Corporon, 2001; Leinert et al., 1997; Pirzkal et
al., 1997; and Li et al., 1994; and using optical spectroscopy:
Corporon \& Lagrange, 1999).

The present study increases the {\it observed} fraction significantly,
presumably because many of the previous studies have lower sensitivity
to fainter companions, and does point toward even larger fractions
when considering observational biases. For example, Pirzkal et al
(1997) found a binary frequency of 23\% from a sample of 39 Herbig
Ae/Be stars.  Taking into account their observational biases they
calculated that the binarity of Herbig Ae/Be stars should be $\sim$
85\%, and find this is comparable to that extrapolated for T Tauri
stars (e.g. Ghez et al., 1993 find an uncorrected binary frequency of
60$\pm$17\%), and exceeds that of near solar type Main Sequence stars
(57\%; Duquennoy \& Mayor, 1991).  These numbers compare well with the
compilation of binary statistics for B main sequence stars by Abt, Gomez \&
Levy (1990), who found an average of 0.7 companions per primary and a
binary frequency, as deduced from their tables, of 55\%.

Finally, the high binary frequency observed in a sample such as the
present one could be the result of observational bias. When dealing
with a magnitude limited sample, one preferentially includes binaries
as they would appear brighter than a single star of the same
type. This bias is larger when the brightness contrast between both
components is smaller. This could be investigated by spatially
resolved imaging at optical wavelengths (yielding the light contrast
directly) or dedicated spectro-astrometry, yielding not only the light
contrast to be measured, but also allowing the splitting of the
spectra, making it possible for the spectral types of the components,
and ultimately the mass ratio, to be determined.
Having said that, the Herbig Ae/Be star catalogue of Th\'e et
al. (1994) is by the subjective nature of which it is compiled,
subject to many biases, amongst which the brightness of the targets is
perhaps the only one that can be quantitatively addressed.

\subsubsection{Binarity as function of spectral type}

Let us now turn to the percentage of binaries within the Herbig Be
stars and Herbig Ae stars.

Among the theories for binary star formation we find the stellar
capture model in small (N$<$ 10) clusters, where the natural outcome
is the larger binary fraction for more massive stars (e.g. McDonald \&
Clarke 1995). Multiple fragmentation models also predict a larger
binary fraction with stellar mass. These consider the breaking up of a
molecular cloud from which fragments, which become the binary
components, can form. The fragmentation models predict a much higher
binary fraction in general than the capture models.  In contrast, the
so-called N=2 fragmentation models are different from the multiple
fragmentation models in the sense that the binary frequency has no
dependence on mass (see e.g. the review by Clarke 2001).

Taken at face value, more binaries were detected around the Herbig Be
stars (11 out of 15; 73$^{+12}_{-16}$\% - interestingly very close to
Pirzkal's prediction) compared with the Herbig Ae stars (4 out of 13;
31$^{+18}_{-14}$\%). Of the 5 possible binary detections 4 have A spectral
types and 1 has an F spectral type. This increases the Herbig Ae
observed binary frequency to 62$^{+15}_{-17}$\%.

The fact that the number of confirmed Herbig Be binaries is larger
than that of the Herbig Ae binaries could point at a real effect.  It
could either mean that Herbig Be stars have a larger percentage of
companions than Herbig Ae stars, or that the spectro-astrometry is
more efficient at detecting companions around Herbig Be stars than
Herbig Ae stars (which is reinforced by noting that all ``possible''
detections were of A or F type, but no B type).


From {\it K}-band imaging Testi et al (1997) found a correlation
between the mass of Herbig Ae/Be stars and the number of {\it K}-band
stars detected around it. From further observations of 44 young stars,
covering almost uniformly the whole spectral range from O9 to A7,
Testi, Palla \& Natta (1999) conclude that rich clusters (with
densities up to 10$^3$ pc$^{-3}$) only appear around stars earlier
than B5. If early B type stars are associated with dense clusters then
one could assume that a higher percentage of Herbig Be stars will have
companions.  Bonnell \& Clarke (1999) show that the observational data
of Testi et al (1999) do not require such an interpretation, but are
instead compatible with the statistics of randomly assembling stars
from the initial mass function into clusters with a range of
sizes. Therefore, a higher binary frequency observed around Herbig Be
stars may not have implications on the initial formation of massive
stars, but may have implications on the further evolution of Herbig Be
stars.
We note that the different statistics between the Herbig Ae and Herbig
Be objects in this paper only become comparable when taking into
account the possible detections.

\begin{table}
\caption{ The binaries that have a measurement of the PA of the
circumprimary disc available.  The third column shows the binary PA,
taken from Table~\ref{binarytable}.  Literature values are denoted
with $^{L}$.  The fourth column shows the intrinsic polarization angle
derived from spectropolarimetry by Vink et al. (2005), the values
between brackets are quoted to be uncertain.  The last
column shows the difference (computed to be in the range 0..90$^{\rm
\circ}$) between the binary PA and the intrinsic polarization PA.
\label{poltable}}
\begin{tabular}{llrrrrrrr}
\hline
 NAME  & Sp Tp  &  PA  &     $\Theta_{\rm Pol}$ & $\Delta$  &  \\
            &      &  (deg)    &  (deg)           &  (deg)  \\
\hline
\hline
 MWC 166  & B0  & 298$^{L}$ & 50     & 68  \\ 
 MWC 361  & B2  & 164$^{L}$ & (90)   & 74  \\
 XY Per   & A2  & 76$^{L}$  & (70)   & 6   \\
 HD 58647  &B9  &  115      &  20    & 85  \\
 MWC 158   &B9  &  30       &  135   & 75 \\
 HD 45677  &B   &  150      &  70    & 80  \\
\hline
\end{tabular}
\end{table}

\subsubsection{On the position angles of the binaries and the circumprimary discs}

An additional question, already alluded to in the Introduction
addresses not only the formation of binary stars, but also the
formation of the most massive Herbig Be stars - for which stellar
mergers have been proposed (see Bonnell, Bate \& Zinnecker 1998). In
the following we investigate whether the derived position angles from
the binaries are of any help in this respect.

A prediction from the above mentioned models concerns the alignment of
the binary systems and the (possibly accretion) discs around the
primary stars. Capture models predict randomly oriented discs around
both components of the binary, while fragmentation models predict
co-planar discs around the stars. For the merger models to still
result in a binary, we need a triple system to begin with.  It is
unlikely that the merger process will result in a disc aligned with
the original tertiary component, hence random orientations would be
expected, as also noted by Bally \& Zinnecker (2005).

Wolf, Stecklum \& Henning (2001) and Jensen et al. (2004) present
studies of T Tauri stars and found the binary systems in their samples
to be preferentially aligned with the discs, thus favouring the
fragmentation model for low mass stars. A similar study was performed
for Herbig Ae/Be stars by Maheswar, Manoj \& Bhatt (2002), who find
that many Herbig binaries are parallel or perpendicular to within
30$^{\circ}$ from the broad band polarization.

We have the sample in hand to improve this sort of studies and extend
it to higher masses. To this end we draw on the spectropolarimetric
results of Vink et al. (2002, 2005).  Inspired by the seminal work by
Poeckert \& Marlborough (1976) who proved the existence of discs
around Be stars with this method, they use the H$\alpha$ line
de-polarization to reveal the presence of small scale electron
scattering discs around pre-main sequence stars. An advantage of this
method is that the intrinsic angle of the polarization can be directly
determined and is independent of the value of the interstellar
polarization.  The latter affects the previous broadband polarization
studies (see e.g. the discussion by Jensen et al. 2004). For optically
thin scattering, which is found to be the case for Herbig Ae/Be stars
(Vink et al. 2005), the intrinsic polarization angle is perpendicular
to the disc.  As the polarization is due to electron-scattering, the
discs are very small (of order stellar radii; Cassinelli, Nordsieck \&
Murison, 1987) and we can be confident that the polarization traces
the disc around the {\it primary} object and not a circumbinary disc.

The sample of objects with a PA determination for both the binary and
the disc measured from the spectropolarimetry contains 7 targets. From
this we have to exclude AB Aur whose data on two occasions resulted in
two different position angles indicating that the binary is too wide
to believe the PA in the first place.  We list the intrinsic
polarization angles of the remaining stars in
Table~\ref{poltable}. The majority of the objects (5 out of 6) have an
angle difference to within $\approx$ 20$^{\rm \circ}$ of
perpendicular, indicating that the circumprimary discs and binaries
are well aligned. XY Per is the exception, with a difference of
6$^{\rm \circ}$.

\begin{figure}
\epsfig{file=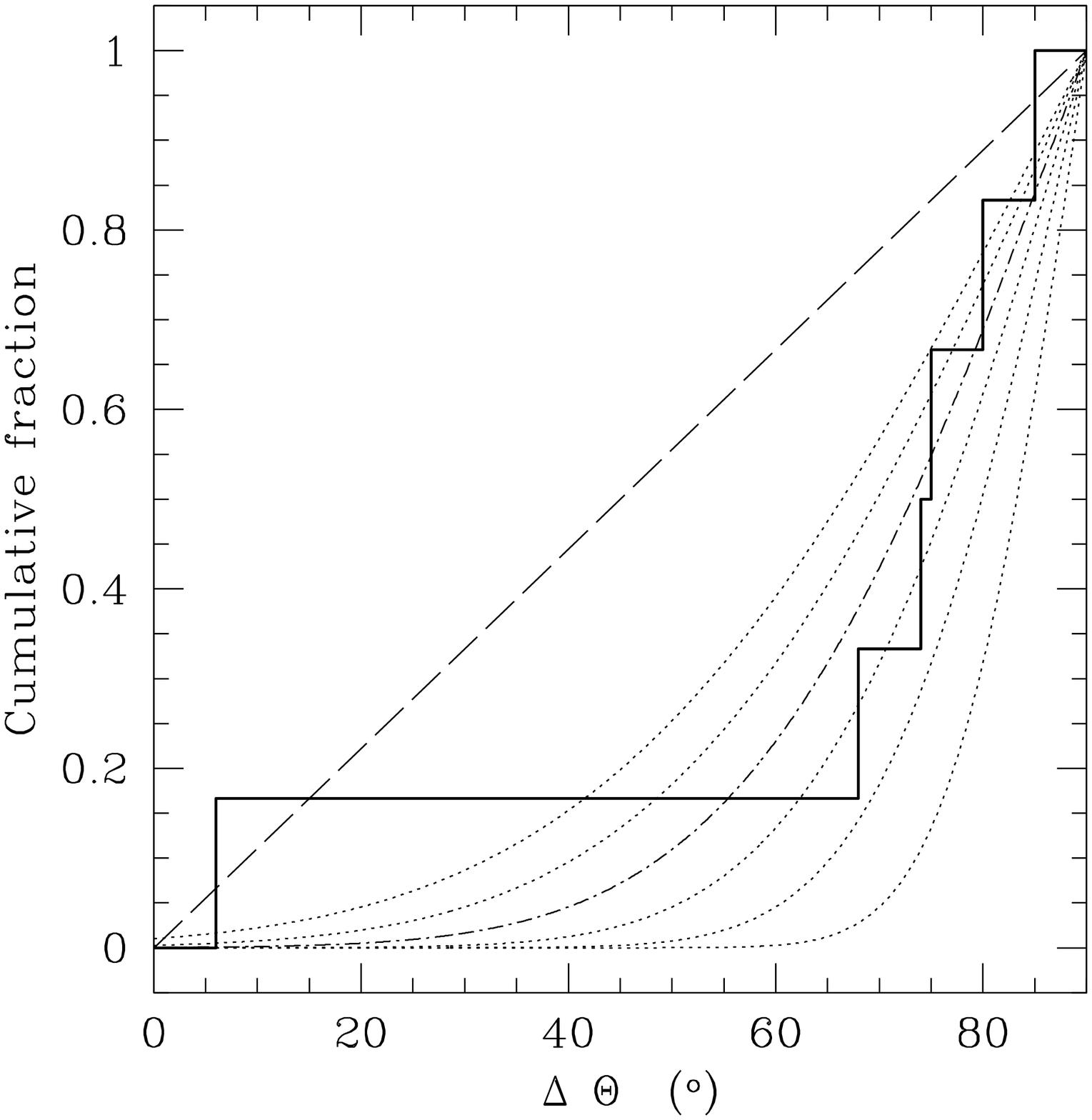, width=0.48\textwidth}
\caption{The cumulative distribution of the differences between the
binary PA and the intrinsic polarization angle. An offset of 90$^{\rm
\circ}$ means that both are aligned (see text). The data, represented
by the solid line, are compared with several distributions. The dashed
line shows the random distribution, while the dotted lines represent
Gaussian distributions with a mean angle difference of 90$^{\rm
\circ}$, and $\sigma$ ranging from (from left to right) 35, 30, 25
(indicated by the dashed-dotted line), 20, 15, and 10$^{\rm \circ}$
respectively. The data are not consistent with a random
distribution. If the discs and binaries are assumed to be aligned, the
data are well matched by a Gaussian distribution with a width
comparable to the observational error.
  \label{KS}}
\end{figure}

To get a quantitative handle on this finding we computed the
cumulative distribution functions of the data and several assumed
distributions, and then calculated the Kolmogorov-Smirnov (KS)
statistic. In Fig.~\ref{KS} we show the cumulative fraction of the
data, of a random distribution of angle differences and of several
Gaussian distributions with an average angle difference of 90$^{\rm
\circ}$.  It can be seen from the figure that the data (represented by
the solid line) are predominately found close to 90$^{\rm \circ}$.
The possibility that the data were drawn from a sample of stars where
the discs and binary systems are randomly aligned is very small.  The
KS statistic returns a probability of 1.8\% that this is the
case. Alternatively, we can reject the models that predict random
orientations at the 98.2\% confidence level. 

We now briefly investigate whether the models predicting aligned discs
and binaries can be rejected with equal confidence or not.  Although
the number of stars under consideration is too small to constrain the
specific shape of the underlying distribution, comparing the data with
some simple models will give an indication of the nature of the
sample.  We computed Gaussian distributions with mean angle
differences of 90$^{\rm \circ}$ and $\sigma$s ranging from 35$^{\rm
\circ}$ to 10$^{\rm \circ}$.  The Gaussians were calculated over the
range 0-180$^{\rm \circ}$ and then reprojected to cover 0-90$^{\rm
\circ}$ as we investigate angle {\it differences}. The results are
also shown in Fig.~\ref{KS}. The broadest and narrowest distributions
do not match the data at all, but the Gaussians with $\sigma$=20$^{\rm
\circ}$ and $\sigma$=25$^{\rm \circ}$, values close to the
experimental error, resemble the data comparatively closely. The KS
test returns probabilities of 89\% and 93\% respectively that the data
could be drawn from such a population.


This provides strong evidence in favour of the fragmentation scenario
that predicts aligned discs and against the stellar capture scenario
which predicts random orientations.  A final remark concerns the
formation of the more massive stars.  The spectral types covered by
this sub-sample of objects extends to B0, tracing masses exceeding the
maximum possible mass that is accumulated in simple accretion models
(e.g. Wolfire \& Cassinelli 1987, Bonnell et al. 1998). It appears
that the stellar merger scenario which predicts random orientations is
also ruled out by our finding.

\section{Conclusions}

We have studied 31 young stars by means of spectro-astrometry, and
find the following.

\begin{enumerate}

\item A binary frequency of 68$\pm$11\% was found in the sample of 28
Herbig Ae/Be stars.  No other technique has been able to find such a high
binary frequency in Herbig Ae/Be stars, demonstrating the proficiency
at which spectro-astrometry can detect binaries.

\item There is a small hint that the Herbig Be stars are more
likely to be found in binaries than the Herbig Ae objects
(73$^{+12}_{-16}$\% versus 31$^{+18}_{-14}$\%, the latter rises to
62$^{+15}_{-17}$\% when the possible detections are included). This
could indicate that Herbig Be stars are more likely to be found in
binaries than their lower mass Herbig Ae counterparts.

\item We  detected asymmetric outflows from Z CMa and HD 87643.  We
conclude that HD 87643 has a high velocity and low velocity outflow,
towards the north. No evidence for jets was present in our data.

\item We present evidence that the circumprimary discs are aligned
with the binary systems. This favours the fragmentation scenario for
the formation of Herbig Ae/Be binaries over stellar capture models
which predict randomly oriented discs. The next step is to identify
the mode of fragmentation, i.e. whether the clouds break up in 2
fragments (predicting roughly fixed mass ratios) or multiple
fragments, predicting random mass functions.

\item The strong alignment, even for early B type stars, provides
evidence against the stellar merger scenario invoked to produce
massive stars. Instead, as these objects have been found to have discs
in the first place, disc accretion remains a viable possibility
(cf. Norberg \& Maeder 2000).

\end{enumerate}

\section*{Acknowledgements}

We thank the referee, Rob Jeffries, for his constructive remarks which
helped improve the paper.  Jorick Vink is thanked for his useful
commments on an earlier version of the manuscript. We wish to thank the
staff of the W\^aldsang in Bakkeveen, where part of this paper was
written, for creating an environment making it conducive to study.  DB
acknowledges support from a student grant from the Particle Physics
and Astronomy Research Council of the United Kingdom which also funded
MP through a post-doctoral grant. The allocation of time on the
Anglo-Australian Telescope was awarded by PATT, the United Kingdom
allocation panel.  Part of the observations are based on data obtained
from the Isaac Newton Telescope, La Palma Spain, in the Spanish
Observatorio del Roque de los Muchachos of the Instituto de Astrof\'\i
sica de Canarias.

\appendix



\section{Spectro-astrometric Simulations of binaries}\label{Simulations}

A spectro-astrometric signature from a binary will arise across a
spectroscopic line if there is a difference in intensity from each
component star. In the case of H$\alpha$ emission from a binary
system, it is highly unlikely that both components have the same
H$\alpha$ intensities and profiles.  Therefore, given the right
conditions a spectro-astrometric signature across H$\alpha$ will be
present. Below, with the aid of simulations, we investigate the
properties of these signatures.

The approach we use is to simulate the spatial profile at each
dispersion pixel, and then build up a 2D spectrum across
H$\alpha$. The spatial profile from a single star at any particular
wavelength can be approximated to a gaussian whose height is in counts
(number of photons), and whose width is in spatial pixels (which
depends on the seeing and spatial resolution of the instrumental
set-up). This is achieved by convolving a point source, where the
seeing determines the standard deviation, $\sigma$, of the gaussian.
In the case of a binary, two point sources at a given separation are
convolved. Since the seeing is identical for each star, the widths of
the  spatial profiles of both components are equal.

The input parameters are the seeing (in arcsecs), the continuum
intensity ratios of the primary and secondary, I$_2$/I$_1$, the
separation {\it d} (in pixels), the size of the spatial pixels (in
arcsec pixel$^{-1}$), and the H$\alpha$ intensity ratio of the primary
and secondary as a function of wavelength, H$\alpha_2$/H$\alpha_1$. In
the simplest of binary systems we assume that only one of the stars is
emitting H$\alpha$ and the other has H$\alpha$ in absorption.  Once
the spatial profiles have been simulated they are treated as real
observations and run through the gaussian fitting program. Both the
position and FWHM spectra are then plotted in the usual
fashion. Experimental errors are not taken into account in these
computations. As an indication of what could be expected, we note that
the high signal-to-noise data such as presented here have rms
variations on the position and FWHM spectra of order
milli-arcseconds. These values are much smaller than the offsets that
are discussed here.

\begin{figure*}
\centering
\epsfig{file=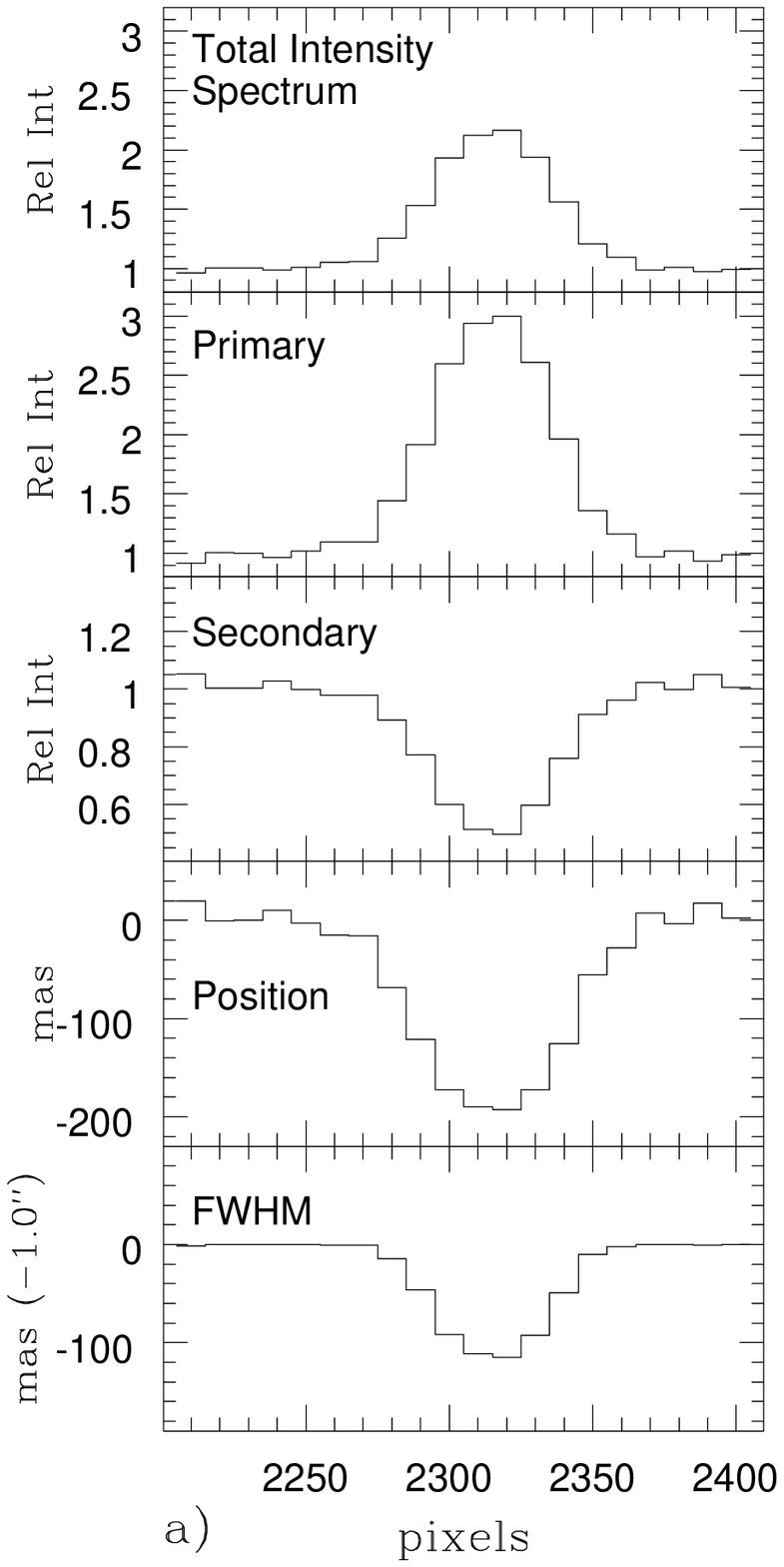,width=4.55cm}
\epsfig{file=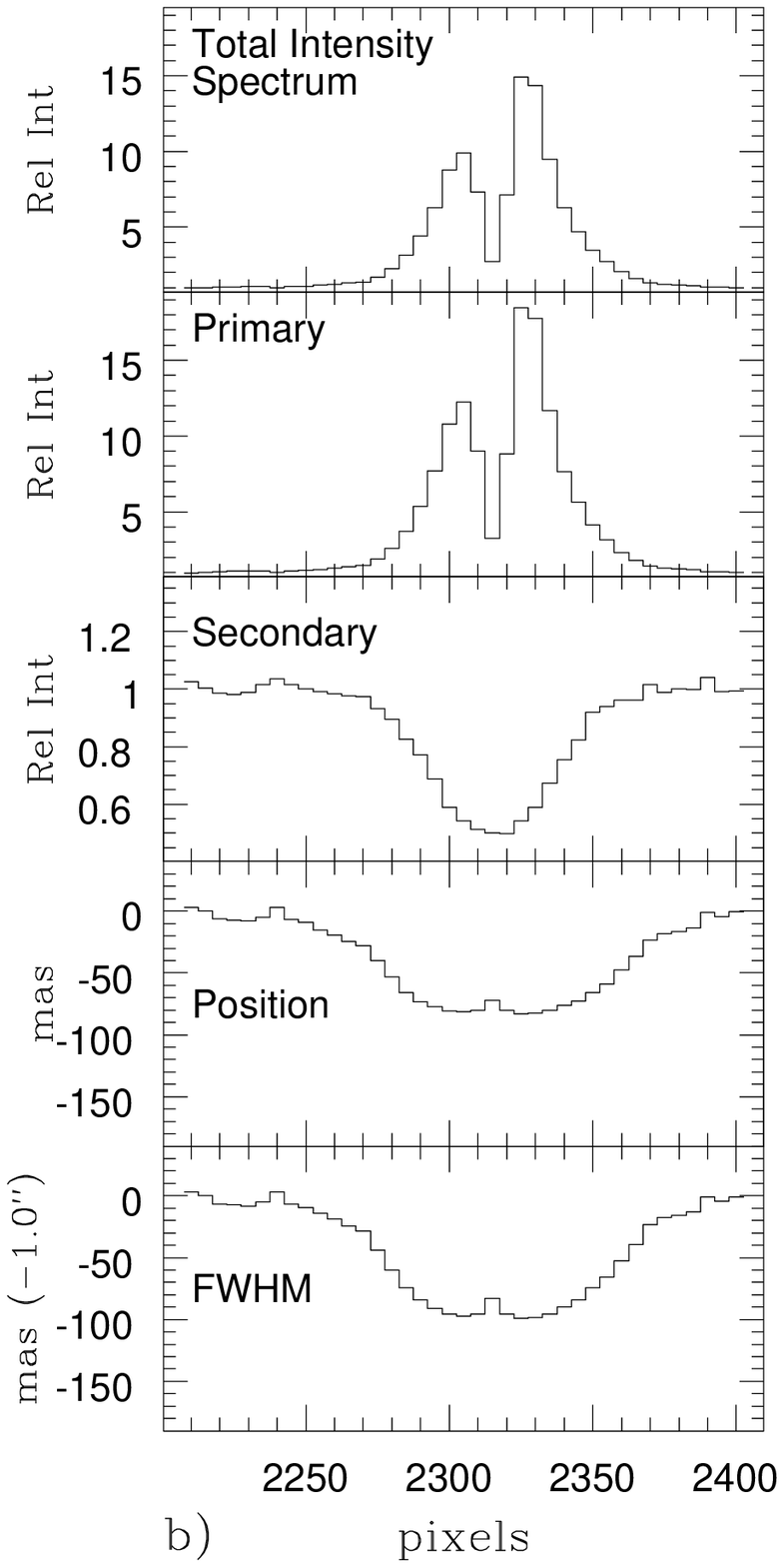,width=4.55cm}
\epsfig{file=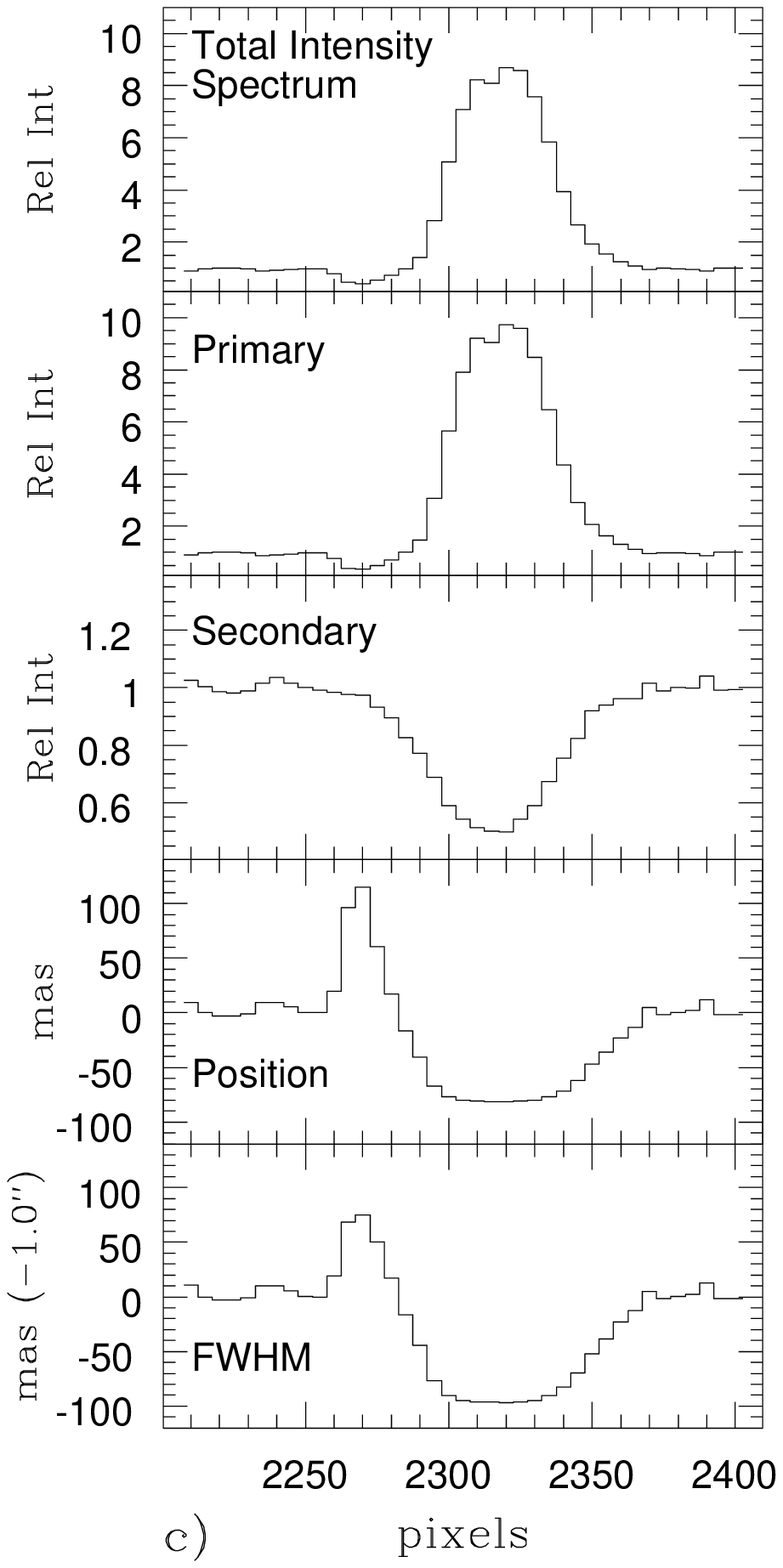,width=4.55cm}
\epsfig{file=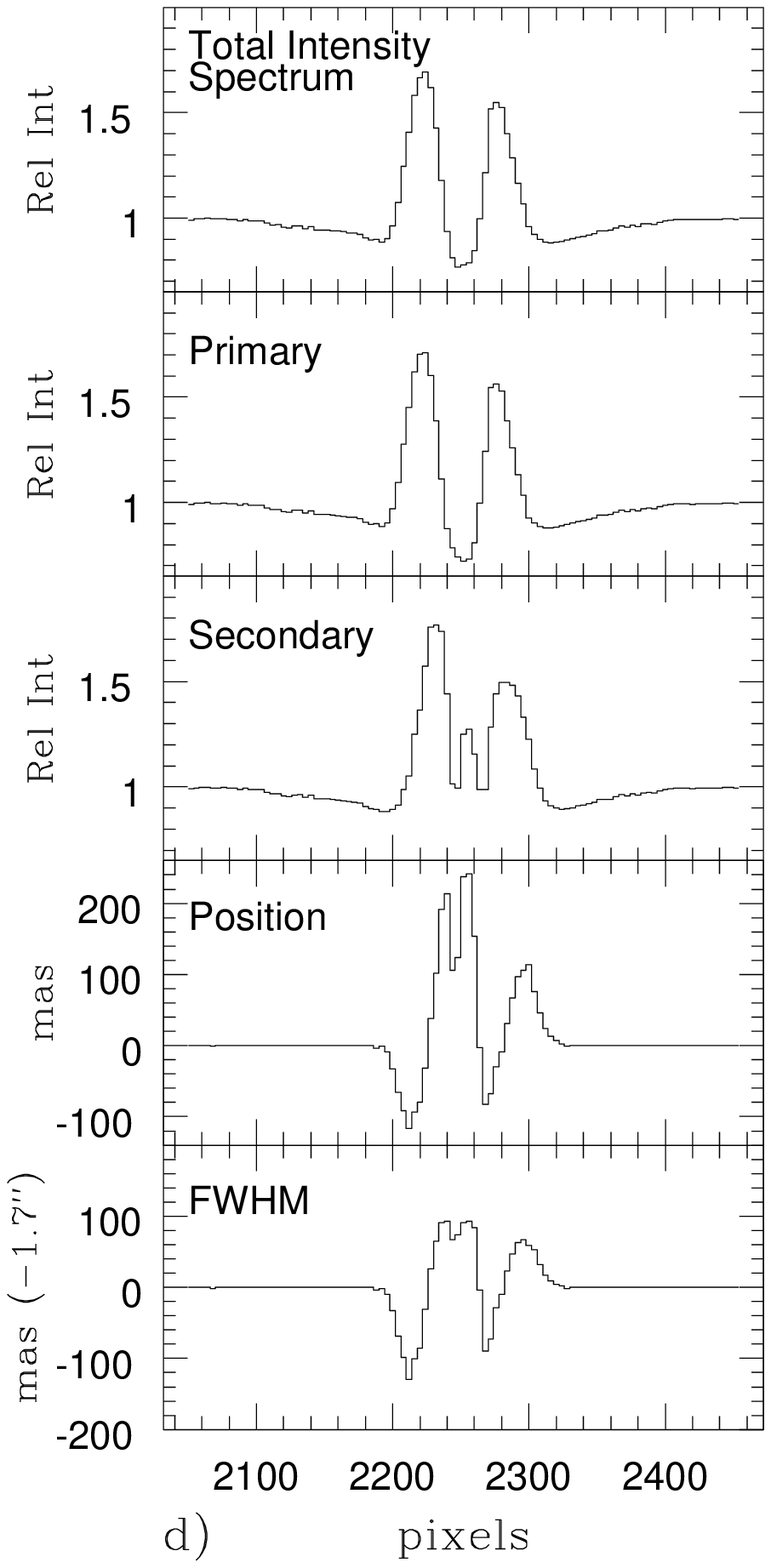,width=4.5cm}\\
\caption{Simulated binary systems. a) The primary has H$\alpha$ in
  emission with a single peaked profile, with seeing = 1.0 arcsec,
  I$_2$/I$_1$ = 1.0, and {\it d} = 0.5 arcsec. b) The primary is
  emitting H$\alpha$ with a double peaked profile, for seeing = 1.0
  arcsec, I$_2$/I$_1$ = 0.5, and {\it d} = 0.5 arcsec. c) The primary
  has a H$\alpha$ P Cygni profile, with seeing = 1.0 arcsec,
  I$_2$/I$_1$ = 1/4, {\it d} = 0.5 arcsec. d) Both primary and
  secondary stars are emitting H$\alpha$. The position and FWHM
  displacements (lower panels) are similar to those of XY Per observed
  on 23 Sept 2002 using the binary parameters of the system, seeing =
  1.7 arcsec, I$_2$/I$_1$ = 0.6, and {\it d} = 1.3
  arcsec. \label{XYPersims}}
\end{figure*}

In the following sections the effects of changing the input parameters
on the spectro-astrometric signatures are investigated.  The system
under consideration is composed of one star with a broad single peaked
H$\alpha$ emission profile, peaking at 3 times the continuum, and the
other star with H$\alpha$ absorption as broad as the emission profile,
and at half the intensity of the continuum. Figure~\ref{XYPersims}
presents the H$\alpha$ profiles of both components for a separation of
0.5 arcsec, seeing of 1.5 arcsec, and I$_2$/I$_1$ = 1. During the
simulations the intensity spectrum of both components is kept fixed,
whilst the intensity spectrum of the system (top panel) varies for
each simulation, as a function of {\it d}, seeing, I$_{2}$/I$_{1}$,
and H$\alpha_2$/H$\alpha_1$.

For all of the following simulations, the size of the spatial and dispersion
pixels are similar to the instrumental set-up at the AAT in January 2002
($\sim$ 0.15 arcsec pixel$^{-1}$ and 0.15 $\rm \AA$ pixel$^{-1}$ (6.7
km \, s$^{-1}$ pixel$^{-1}$) respectively, see section \ref{Observations}).

\subsection{Changing the separation}\label{SEPsec}

\begin{figure}
\centering
\epsfig{file=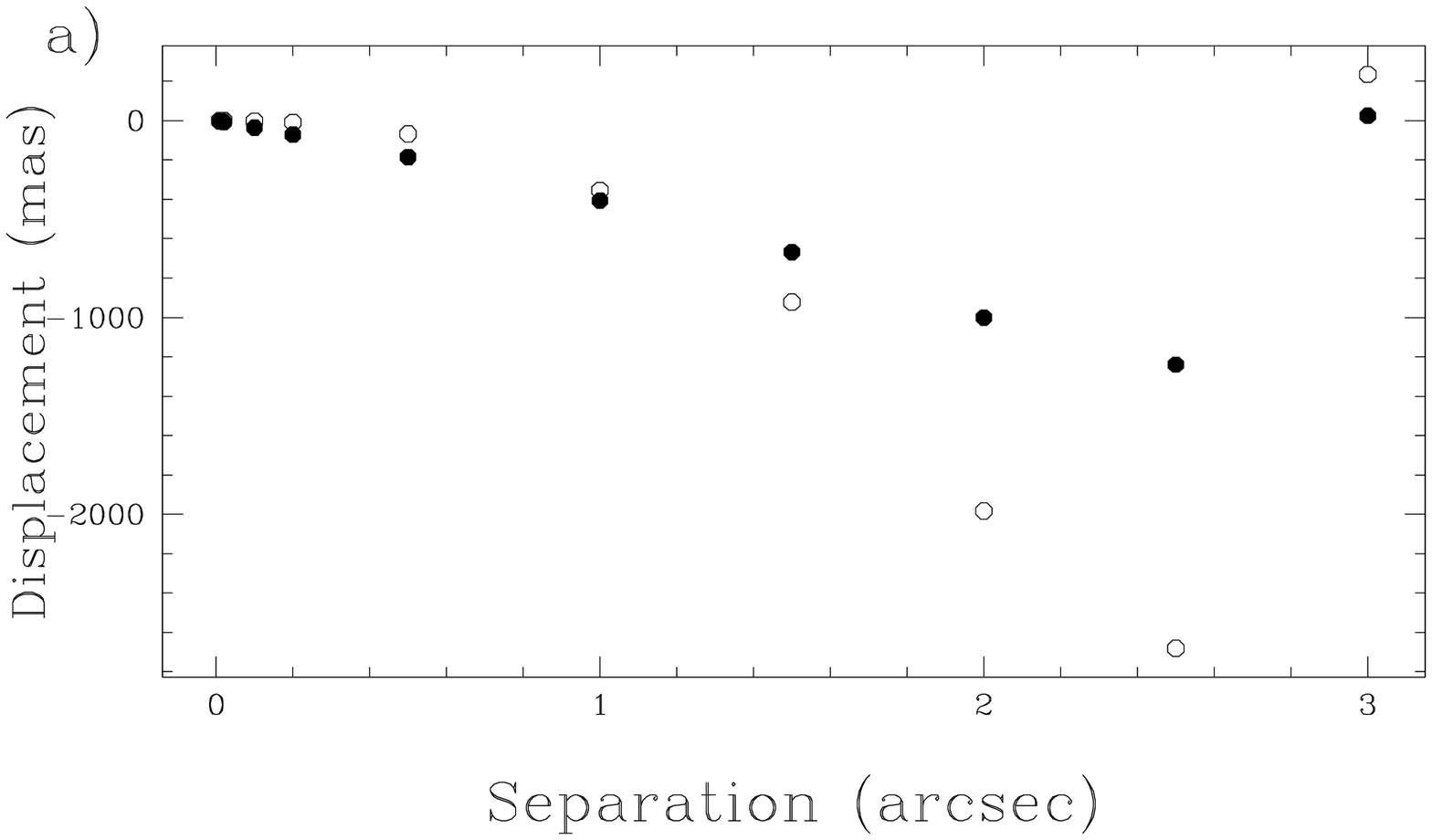,width=7.8cm}\\
\epsfig{file=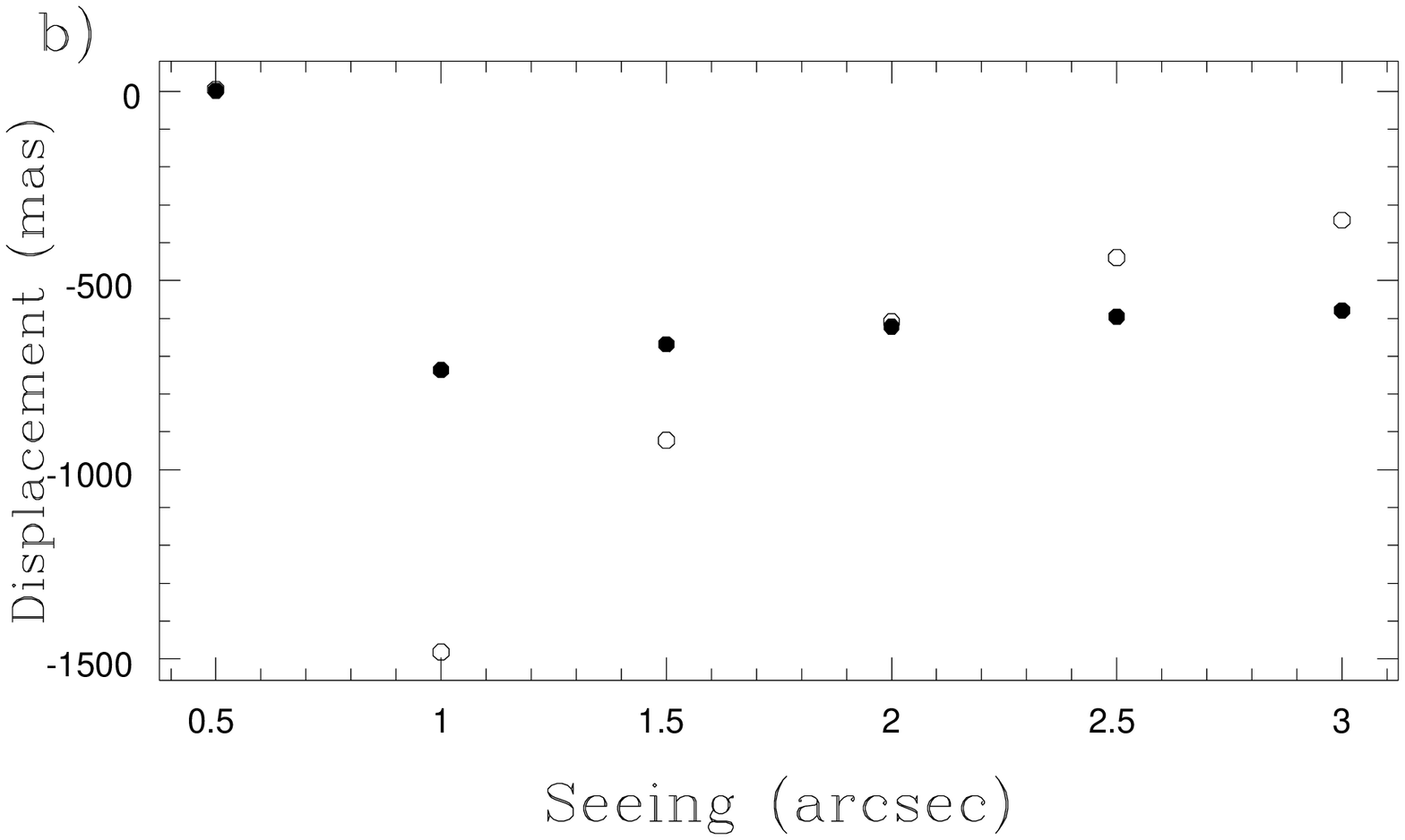,width=7.8cm}\\
\epsfig{file=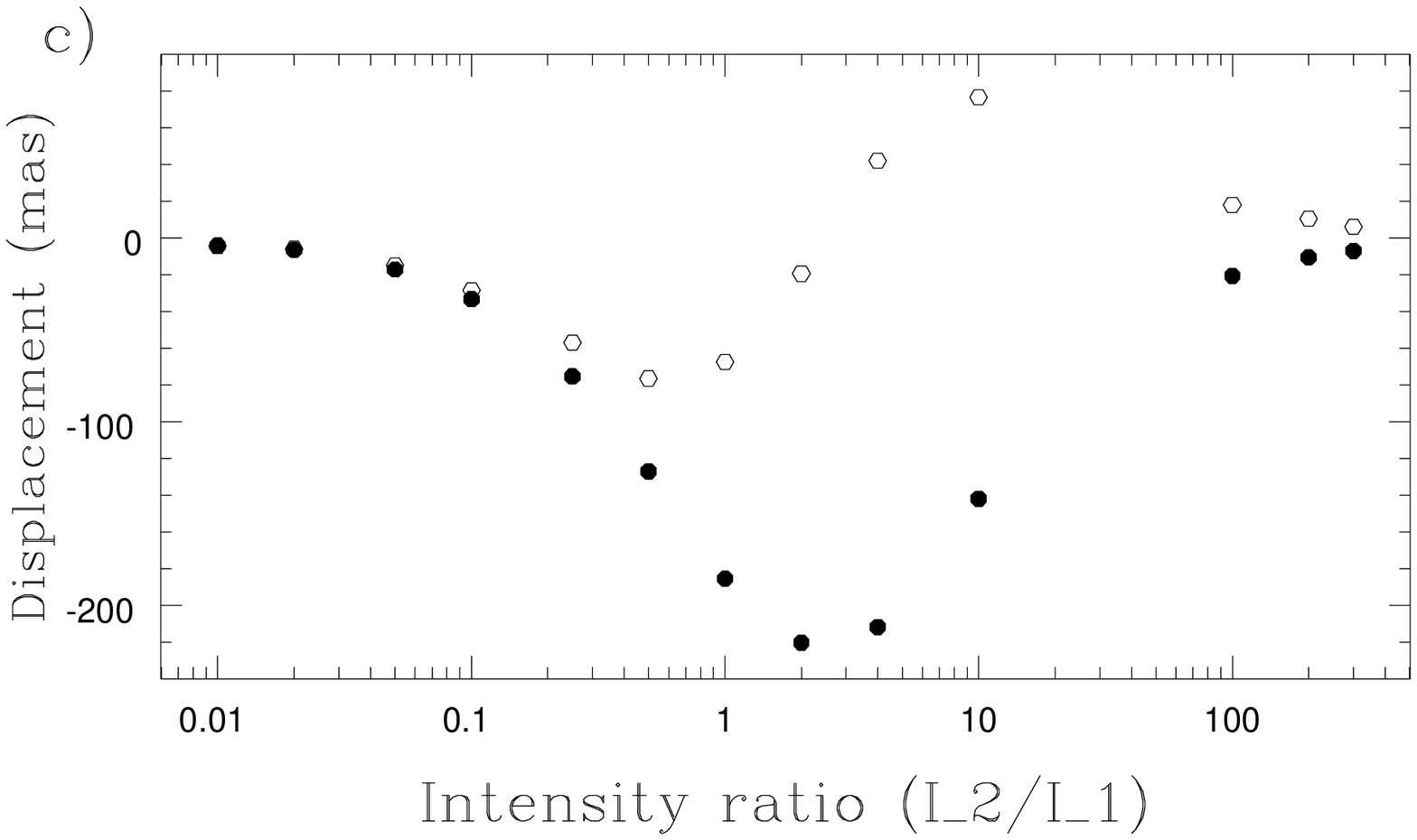,width=7.8cm}\\
\epsfig{file=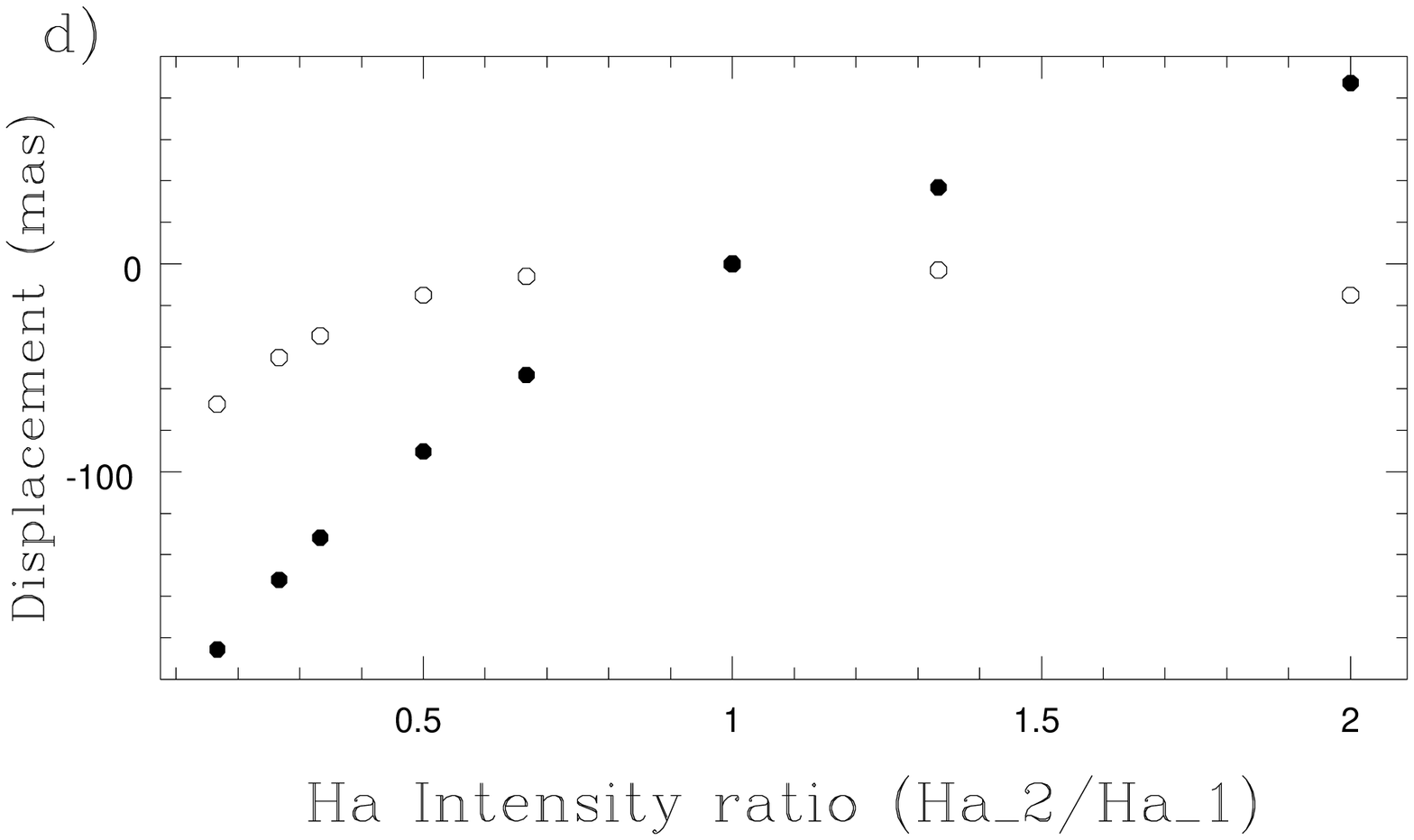,width=7.8cm}\\
\caption{The maximum positional displacements (closed circles) and
FWHM displacements (open circles) against the parameters binary
separation, seeing, binary continuum intensity ratio, and binary
H$\alpha$ intensity ratio. For panel a) I$_2$/I$_1$ = 1 and seeing =
1.5 arcsec; panel b) I$_2$/I$_1$ = 1 and {\it d} = 1.5 arcsec; panel
c) {\it d} = 0.5 arcsec and seeing = 1.5 arcsec; panel c) {\it d} =
0.5 arcsec, I$_2$/I$_1$ = 1 and seeing = 1.5 arcsec. 
\label{simulations}}
\end{figure}

Here we change the separation of the binary system whilst all other
input parameters are kept constant. A seeing of 1.5 arcsec,
representative for the observations discussed in this paper, and
continuum intensity ratio, I$_2$/I$_1$ = 1, is chosen. Figure
\ref{simulations}a shows the maximum positional and FWHM displacement
at H$\alpha$ from the continuum for several binary separations. For
clarity the position and FWHM spectra are centred on zero at the
continuum.

As {\it d} increases, the position of the continuum moves away from
the H$\alpha$ emitting star simply because for I$_2$/I$_1$ = 1 the
continuum is located at {\it d}/2. As we move along $\lambda$ towards
the H$\alpha$ maximum, the positional displacement moves towards the
dominant H$\alpha$ emitting star. This displacement becomes larger
with larger separations.

An increase in the positional displacement as the separation is
increased will halt when the spatial profile of the two components
becomes fully resolved. This occurs in figure \ref{simulations}a at
{\it d} around 3.0 arcsec. At wide separations the technique of
spectro-astrometry breaks down because only one profile (in this case
the primary star) is fitted by the fitting program at both the
continuum and H$\alpha$ maximum, i.e. spectro-astrometry should only
be applied to unresolved binaries. For a separation of 3.0 arcsec the
FWHM increases because the spatial profile wings of the primary begin
to overlap with the secondary and the fitting program again includes
some of the secondary.

For this simulated system a displacement of 0.03 pixels is visible for
a separation of only 10 mas. This is a shift of 4.5 mas, which for
many of our observed stars is better than the spatial sensitivity, and
would therefore be detectable if the observing conditions are similar
to, or better than, the simulated parameters. For separations $<$ 1.0
arcsec (the slit width) the positional displacement is larger than the
FWHM displacement, at 1.0 arcsec the positional and FWHM displacements
are almost identical, and at separations $>$ 1.0 arcsec the positional
displacement $<$ FWHM displacement.

\subsection{Changing the seeing}\label{SEEsec}

Figure \ref{simulations}b displays the maximum spectro-astrometric
displacements for {\it d} = 1.5 arcsec, I$_{2}$/I$_{1}$ = 1, and varying
seeing (from 0.5 to 3.0 arcsec). A decrease in the seeing increases
both the positional and FWHM displacement. In this case, for a seeing
of 2.0 arcsec and larger, the positional displacement is larger than
the FWHM displacement. But for a seeing between 2.0 and 0.5 arcsec the
FWHM displacement becomes larger than the positional displacement and
the effect of the seeing is rather more significant in the FWHM
displacements than in the positional displacements. It is therefore
important to approximately know the intrinsic seeing before
interpreting the spectro-astrometric signatures, particularly since
the seeing can vary considerably between observations. The
displacements at a seeing of 0.5 arcsec are zero because the binary is
resolved.

\subsection{Changing the intensity ratios}\label{INTsec}

Figure \ref{simulations}c show the maximum spectro-astrometric
displacements when varying the continuum intensity ratios,
I$_2$/I$_1$, whilst keeping the separation (0.5 arcsec) and seeing
(1.5 arcsec) constant. I$_1$ is the H$\alpha$ emitting star, whilst
I$_2$ has H$\alpha$ in absorption. Interestingly, one still sees a
positional and FWHM displacement across H$\alpha$, of 4.5 mas, when
the secondary star is a 100th the intensity of the primary star
(i.e. a difference of 5 mag). As I$_2$/I$_1$ increases to 1, the
positional and FWHM displacements increase, and for this separation
(0.5 arcsec), the positional displacements increase at larger
increments than the FWHM displacements.

At I$_2$/I$_1$ $>$ 1 the H$\alpha$ emitter is now the
secondary star, and the primary has H$\alpha$ in absorption. As
I$_2$/I$_1$ increases (from 2 upwards), the positional displacements
decrease until no displacement is present, when the H$\alpha$ emitting
star is too faint to contribute to the total H$\alpha$ intensity of
the system. This is found to occur at I$_2$/I$_1$ $>$ 300 for this
system. At this point the total H$\alpha$ profile approximates to the
H$\alpha$ profile of I$_2$. A displacement of 6 mas in both the
position and FWHM spectrum, is present when I$_2$/I$_1$ is as large as
300 (a difference of $\sim$ 6.2 mag). In this case no H$\alpha$
emission is present in the total intensity spectrum, but the
spectro-astrometry still detects the binary.

The FWHM maximum displacements are largest when the primary and
secondary star have similar intensities. When the primary is the
H$\alpha$ emitter (I$_2$/I$_1$ $\leq$ 1) the FWHM decreases across
H$\alpha$, whereas when the secondary is the H$\alpha$ emitter the
FWHM increases.

\subsection{Changing the H$\alpha$ intensity ratios}

Here we investigate the effect of changing the H$\alpha$ intensity
ratios, H$\alpha_2$/H$\alpha_1$, whilst the seeing (1.5 arcsec), {\it
d} (0.5 arcsec), and I$_2$/I$_1$ (1) are kept constant. One component
spectrum is kept constant, with a single peaked H$\alpha$ emission
profile, peaking at 3 times the continuum (identical to the primary
spectrum of figure \ref{XYPersims}a). The other component H$\alpha$
profile is varied.

The results are presented in figure \ref{simulations}d. At
H$\alpha_2$/H$\alpha_1$ = 1 (and in this case I$_2$/I$_1$ = 1), the
two component H$\alpha$ profiles are identical, therefore we see no
displacements in either the position or FWHM spectrum. As the value of
H$\alpha_2$/H$\alpha_1$ moves further from 1 the positional and FWHM
displacements become larger. In the case of the position spectrum, the
displacements lie on either side of the position continuum, depending
on which star is dominating (i.e. for H$\alpha_2$/H$\alpha_1$ $>$ 1,
and H$\alpha_2$/H$\alpha_1$ $<$ 1). In the case of the FWHM spectrum
(bottom panel), any value of H$\alpha_2$/H$\alpha_1$ $\neq$ 1 will
result in a decrease in the FWHM.

\subsection{Changing the H$\alpha$ profiles}\label{Haprofiles}

In the above simulations the H$\alpha$ emission profile was kept as a
simple single peaked profile. However many Herbig Ae/Be stars have a
double peaked or P Cygni H$\alpha$ profile (see for example Reipurth
et al., 1996). Figures \ref{XYPersims}b and c show examples of
spectro-astrometric displacements when the primary star of a binary
system has a double peaked H$\alpha$ emission profile (from MWC 158)
or a P Cygni H$\alpha$ emission profile (from AB Aur). In both cases
the secondary star has been chosen to have an H$\alpha$ absorption
profile with maximum absorption at 0.5. For figure \ref{XYPersims}b,
the position and FWHM displacements return towards the continuum
across the H$\alpha$ minimum because the intensity of the primary
diminishes. This produces a 'double peaked' profile in both the
position and FWHM spectra. The lower panels of figure \ref{XYPersims}c
shows an example of the position and FWHM spectra when the primary has
a P Cygni H$\alpha$ profile. At the peak in the emission we see a
typical binary signature, where there is a decrease in the FWHM and a
displacement towards the emitting star in the position spectrum. At
the P Cygni absorption there is a displacement towards the secondary
star as the intensity of the primary decreases. When I$_2$/I$_1$ $<$ 1
the FWHM increases across the P Cygni absorption, again because the
intensity of the primary decreases and the secondary becomes more
dominant than at the continuum.

In figure \ref{XYPersims}d both primary and secondary stars are
emitting H$\alpha$. Since XY Per may be in a system with both
components emitting H$\alpha$ we choose to use binary parameters
similar to XY Per, i.e. {\it d} = 1.3 arcsec, I$_2$/I$_1$ = 0.6 , and
seeing = 1.7 arcsec. The positional and FWHM displacements are similar
to those of 23rd September 2002 of XY Per in the EW direction. The
triple peaked H$\alpha$ profile of the secondary is not apparent in
the total intensity spectrum, but does affect the positional and FWHM
displacements.

\subsection{Summary}

These simulations show how the spectro-astrometric signature from a
binary is dependent on the separation of the component stars, the
continuum intensity ratio of the two stars, the H$\alpha$ profiles,
and the seeing. The main conclusions from these simulations are: a)
the technique is capable of detecting binaries between separations of
$\sim$ 0.01 - 3.0 arcsecs, up to a $\Delta$mag of 5 (and as large as
$\Delta$mag = 6 if the secondary star dominates the H$\alpha$
emission), for reasonable seeing values ($\sim$ 1.5 arcsec); b) the
size of a positional displacement compared with that of a FWHM
displacement is an important indication of the binary separation,
i.e. binaries with separations approximately $>$ the slit width will
have FWHM displacements as large as, or larger than, the positional
displacements.  As outlined by Porter et al. (2004), dedicated
observations and modelling are needed to retrieve the separation and
the individual spectra of both components.

\end{document}